\documentclass[pra,aps,amssymb,nofootinbib,twocolumn]{revtex4}
\usepackage{amsmath}
\usepackage{amssymb}
\usepackage{amsthm}
\usepackage{amsfonts}
\usepackage{algorithmic}
\usepackage{enumerate}
\usepackage{latexsym}
\usepackage[dvips]{graphicx}

\newcommand{\beq}{\begin{equation}}
\newcommand{\eneq}{\end{equation}}
\newcommand{\beqs}{\begin{equation*}}
\newcommand{\eneqs}{\end{equation*}}

\input{epsf}

\begin{document}

\tolerance 10000

\title{On Quantum Bosonic Solids and Bosonic Superfluids}

\author { Zaira Nazario$^\dagger$ and
David I. Santiago$^{\dagger, \star}$ }

\affiliation{ $\dagger$ Department of Physics, Stanford
University,Stanford, California 94305 \\ 
$\star$ Gravity Probe B Relativity Mission, Stanford, California 94305}

\begin{abstract}
\begin{center}

\parbox{14cm}{We review the nature of superfluid ground states and the
universality of their properties with emphasis to Bose Einstein
Condensate systems in atomic physics. We then study the superfluid
Mott transition in such systems. We find that there could be two types
of Mott transitions and phases. One of them was described long ago and
corresponds to suppression of Josephson tunneling within superfluids
sitting at each well. On the other hand, the conditions of optical
lattice BEC experiments are such that either the coherence length is
longer than the interwell separation, or there is too small a number
of bosons per well. This vitiates the existence of a superfluid order
parameter within a well, and therefore of Josephson tunneling between
wells. Under such conditions, there is a transition to a Mott phase
which corresponds to suppression of individual boson tunneling among
wells. This last transition is in general discontinuous and can happen
for incommensurate values of bosons per site. If the coherence length
is small enough and the number of bosons per site large enough, the
transition studied in the earlier work will happen.}

\end{center}
\end{abstract}

\date{\today}

\maketitle
\section{Introduction}

We have recently found \cite{us} that bosonic systems in optical
lattices can have {\it discontinuous transitions} from a superfluid
phase to a Mott insulating phase irrespective of commensurability of
the number of bosons and the number of lattice sites. This is contrary
to expectations from the early theoretical work \cite{fisher}. The
discrepancy is not due to incorrectness in the early work, but due to that,
depending on the physical conditions, the nature of the transition and
the insulating phase can be different. Specifically for a small number
of bosons per site and/or for superfluids with coherence length larger
than the well size, the transition studied in the early work is
impossible \cite{us2}, as it presupposes a superfluid order parameter
within each well.

The nature of the Mott phase of the early work \cite{fisher} is that of
wells of superfluid, where the large repulsion prevents Josephson
tunneling much as it happens in Josephson junctions arrays when the
charging energy becomes large. The nature of our phase corresponds to
when localization of single atoms in the wells destroys the superfluid
order parameter. Our transition is necessarily first order as the
superfluid order parameter, and hence the superfluid response, jumps
discontinuously at the transition. The reason our transition
necessarily destroys the superfluid order parameter is that, since our
transition happens when the coherence length is larger than the well
size or the interwell separation, boson localization introduces a
length scale, the lattice constant, which is
smaller than the coherence length for our transition. This leads to
an energy scale larger than the superfluid ordering scale.

A Mott phase transition has recently been observed
experimentally \cite{ib} but the nature of the transition and of the
Mott phase were not determined from such experiments. On the other
hand, recent experiments are consistent with a discontinuous
transition irrespective of commensurability \cite{ari}.  

In the present article we present the details leading to the
conclusions of our earlier work. We present for the first time the
explicit calculations, and we expand on the nature of the physics that
controls when our transition is expected to happen as opposed to the
transition predicted in the early work \cite{fisher} is expected to
take place. Since issues of superfluid ordering and the different
energy scales in Bose Einstein Condensates (BECs) are essential for a
microscopic understanding we start the article with sections that
review well known physics of BECs and superfluids and their different
length scales. We then study the nature of tunneling in optical
lattices to review old results of ours \cite{us2} on the nature of
tunneling in lattices of bosonic systems. When Josephson tunneling is
suppressed before superfluid correlations are destroyed, the well
known transition \cite{fisher} follows. When superfluid correlations
are destroyed before Josephson tunneling is suppressed, the transition
we recently discovered follows \cite{us}. We then move on to estimate
the properties of the Mott phase and the energetics of when the Mott
phase wins over the superfluid phase leading to a localization
transition.

\section{Simple Bosonic Fluid Systems}

In the present section we will review for completeness well known
results in BECs and superfluid bosonic systems. Most of this knowledge
is due to studies of superfluid He$^4$ \cite{pines} which predates the
experimental realization of artificially engineered BECs
\cite{beca,becb,becc}. These are superfluid systems as they posses a
finite sound speed \cite{bec2,landau,bog} and suppressed
long-wavelength scattering \cite{bec3,bec4}. The superfluid phase has
off-diagonal order with breaking of gauge invariance \cite{phil},
i.e. $U(1)$ phase invariance, characterized by a coherent ground state
of Bogolyubov pairs \cite{bog}.

Even though much of the material in this section is far from new, we
included it in the article for completeness and perspective for the
subsequent sections. We also hope to give a flavor and emphasize the
ideas of macroscopic exactness that characterize stable quantum and
thermodynamic phases of matter. While this perspective is not new, it
is hard to tell if its importance is properly recognized in the BEC
and atomic physics community. It is our wish to stress its importance.

\subsection{Noninteracting Boson Gas}

A gas of noninteracting bosons is described by the Hamiltonian
\beq
\mathcal{H} = \sum_{\vec k} \epsilon_{\vec k}a_{\vec k}^\dagger a_{\vec k} 
\eneq
\noindent where the operators $a_{\vec k}$ and $a_{\vec k}^\dagger$ follow the 
harmonic oscillator commutator relations 
\beq
[ a_{\vec k}^\dagger, a_{\vec k'}^\dagger ] = [ a_{\vec k}, a_{\vec k'} ] = 0 
\; ; \quad [ a_{\vec k}, a_{\vec k'}^\dagger ] = \delta_{{\vec k}, {\vec k'}}
\eneq
\noindent In the continuum limit, the sum becomes and integral and 
$\epsilon_0 = 0$. The dispersion relation is 
\beq
\epsilon_{\vec k} = \frac{ \hbar^2 k^2 }{ 2 m }
\eneq
\noindent While in the continuum limit $| {\vec k}_{min} = 0 |$, in a finite 
box of size L with periodic boundary conditions 
\beq
{\vec k} = \frac{ n_x \pi }{ L } \hat{ x } + \frac{ n_y \pi }{ L } \hat{ y } + 
\frac{ n_z \pi }{ L } \hat{ z }
\eneq
\noindent where $n_x$, $n_y$, and $n_z$ are integers different from
0. Thus in the box $ | {\vec k}_{min} | = \frac{ \pi \sqrt{ 3 } }{ L }
$ and there is a gap that vanishes in the thermodynamic limit, that
is, it vanishes as the number of bosons goes to infinity such that the
density is fixed. For an optical lattice system the boson dispersion
at long wavelengths is exactly as the one above for an appropriately
chosen mass.

The ground state wavefunction is trivial.
\beq
|\Psi_0 \rangle = a_0^N | 0 \rangle
\eneq
\noindent where $a_0 = a_{{\vec k}_{min}}$. This wavefunction reflects
the fact that bosons can occupy the same state, so the ground state
will consists of all N particles in the minimum energy state. This
state has energy $\epsilon_0 = 0$ in the continuum limit. All the
particles are coherent and have the ``same phase'': Bose Einstein
condensation has occurred. This is the preamble to the occurrence of
the phenomenon of superfluidity. Bose condensation is a necessary
condition for superfluidity to occur, but it is not a sufficient
condition. Bose condensation implies that, when the continuum limit is
an appropriate approximation, the system is a quantum fluid. As
explained below, a noninteracting Bose condensate lacks the necessary
rigidity of the ground state wavefunction. A rigid ground state
wavefunction means that the dispersion relation is such that as ${\bf
k} \rightarrow 0$, the excitation of quasiparticles costs an energy
higher than $\frac{ \hbar^2 {\vec k }^2 }{ 2 m }$. Dissipation processes
are prohibited when the system has enough rigidity as they would
violate energy momentum conservation. Dissipationless response is the
definition of superfluidity as the coherence of the condensate can
then survive weak enough perturbations. This necessary rigidity is
provided by repulsive interactions among the bosons comprising the
superfluid. In the absence of such interactions, the lack of rigidity
permits scattering processes to occur which degrade coherence no
matter how weakly the system is perturbed.

\subsection{Interacting Boson Gas}

BECs are really systems of bosons with repulsive interactions among
them. They are supersaturated quantum vapors because they are
metastable. They behave like quantum mechanical ground states for some
time, but due to interactions or trap effects they loose stability and decay. 
Since they live long enough to do measurements of their properties, and we
can do calculations that agree with experiments as if they were ground
states, it is not incorrect to treat them as ground states for times
longer than their lifetimes. The finite lifetimes are both due to
increasing interactions and trap effects. Usually their lifetimes are 
``long enough'' that a lot of ``ground state'' physics can be studied. We 
concentrate in this latter physics.

We will look here at some of the properties of BECs. Another system
that can be described in an exactly the same manner and is actually a
ground state is superfluid He$^4$. The fact that is a liquid is not
contradictory since the elementary excitations of the liquid have to
repel each other as a stability condition. Otherwise, the elementary
excitations can condense leading to some other phase of matter.

We now proceed to describe the ground state and low energy excitation
properties of the interacting Bose gas. We will concentrate on a boson
system on a lattice with onsite Hubbard repulsion as we have bosons
systems in optical lattices in mind. We do emphasize that the physics
we describe here is universal as long as the system is a quantum
fluid, i.e. there exists a macroscopically occupied lowest momentum
state, and there is repulsion among the long wavelength elementary
excitations of the system. When these two conditions are met, the
physics is independent of the microscopic details of your system. 

We now concentrate on the Bose Hubbard Hamiltonian in a lattice (which
of course can be an optical lattice)
\begin{multline} \label{hami}
\mathcal H = \frac{-t}{2} \sum_{ <ij> } ( a_i^\dagger a_j + a_j^\dagger a_i ) 
- \mu \sum_i a_i^\dagger a_i \\
+ U \sum_i a_i^\dagger a_i ( a_i^\dagger a_i - 1 ) - \varepsilon \sum_i 
a_i^\dagger a_i
\end{multline}
\noindent where $\mu$ is a chemical potential, $U$ is the repulsion
between bosons, $t$ is the tunneling amplitude between sites or
hopping term, and $\varepsilon$ represents the depth of the potential
well. In a Mott insulating phase with localized bosons $\mu$ is
irrelevant and can be set equal to 0. It is needed to describe the
superfluid phase, as the approximations made in the Bogolyubov
Hamiltonian make the number of particles to not be conserved. In that
case $\mu$ is chosen in such a way that the average number of
particles equals the total number of particles in the system. In the
thermodynamic limit, the fluctuations around the average number of
particles become negligible compared to the average. We emphasize that
the nonconservation of particles in the Bogolyubov approximation is an
artifact of the approximation. Physically the total number of bosons in the
system is, of course, constant, but the number of bosons in the
condensate cannot be constant due to spontaneous breaking of phase
invariance\cite{phil}.

We consider a system with $M$ bosons and $N$ lattice sites. The number of 
bosons need not be commensurate with the lattice, i.e. $M/N$ need not be an 
integer. The Hubbard interaction term in the Hamiltonian is after Fourier 
transforming
\begin{multline}
\mathcal{H}_I = \frac{U}{N} \sum_{ {\vec q}, {\vec k}_1, {\vec k}_2 } 
\sum_{ {\vec k}_3, {\vec k}_4 } a_{{\vec k}_3}^\dagger a_{{\vec k}_1} 
a_{{\vec k}_4}^\dagger a_{{\vec k}_2} \\
\times \delta ( {\vec q} + {\vec k}_1 
- {\vec k}_3 ) \; \delta ( {\vec k}_2 - {\vec k}_4 - {\vec q} )
\end{multline}
\noindent Performing some momentum summations we get to

\beq
\mathcal{H}_I = \frac{U}{N} \sum_{ {\vec k}_1, {\vec k}_2 } \sum_{\vec q}
a_{ {\vec k}_1 + {\vec q} }^\dagger a_{{\vec k}_1} 
a_{ {\vec k}_2 - {\vec q} }^\dagger a_{{\vec k}_2} 
\eneq
\noindent The kinetic energy is a sum over pairs of nearest neighbors
which describes tunneling from site to site:

\beq\label{Hk}
\mathcal {H}_K = \frac{-t}{2} \sum_{ \delta, i } ( a_{i +
\delta}^\dagger a_i + a_i^\dagger a_{i + \delta} ) = -t \sum_{ \delta,
i } a_{i + \delta}^\dagger a_i
\eneq
\noindent where $\delta$ is one of the vectors from a particle to its
nearest neighbor. Fourier transforming diagonalizes the kinetic energy
giving
\begin{align} \label{Hk2}
\nonumber \mathcal {H}_k &= -t \sum_{ \delta, {\vec k} } e^{ - i {\vec
k} \cdot {\bf r}_\delta } a_{\vec k}^\dagger a_{\vec k} \\
&= \sum_{\vec k} \epsilon_{\vec k} n_{\vec k} 
\end{align}
\noindent where $\epsilon_{\vec k}$ is the dispersion
\beq
\epsilon_{\vec k} = -t \sum_{\delta} e^{ - i {\vec k} \cdot {\bf r}_\delta }
\eneq
\noindent for a cubic lattice, with the energy measured from the state
of zero momentum
\beq \label{epscub}
\epsilon_{\vec k} = - 2 t ( \cos{ k_x a } + \cos{ k_y a } + \cos{ k_z a } ) 
+ 6 t
\eneq
\noindent This shift of the zero of energy is irrelevant to the
physics as it can be absorbed in a redefinition of the chemical
potential. The dispersion goes like $t a^2 k^2 = \hbar^2 k^2 / 2 m$ in
the long wavelength limit. We see that efficient tunneling corresponds
to light particles and suppressed tunneling to very heavy ones. The
full interacting boson Hamiltonian is then
\begin{multline} \label{fullham}
\mathcal{H} = \frac{U}{N} \sum_{ {\vec k}_1, {\vec k}_2 } \sum_{\vec q}
a_{ {\vec k}_1 + {\vec q} }^\dagger a_{{\vec k}_1} 
a_{ {\vec k}_2 - {\vec q} }^\dagger a_{{\vec k}_2} \\
+ \sum_{\vec k} \epsilon_{\vec k} a_{\vec k}^\dagger a_{\vec k} 
- (\mu + \varepsilon + U) \sum_{\vec k} a_{\vec k}^\dagger a_{\vec k} \\
- 6 t \sum_{\vec k} a_{\vec k}^\dagger a_{\vec k} \qquad \qquad \qquad \qquad
\end{multline}

Now that we have ended with the Bose Hubbard Hamiltonian, we will consider 
ground state and low energy excitation properties. At low temperatures 
a macroscopic number of bosons occupy the zero momentum state 
{\it as long as $U$ is not too high}. This is what it 
means to be Bose condensed. We take the number of bosons in the condensate to 
be $M_0$. A quantum state with a large quantum number 
behaves classically. An operator acting on that state corresponding to the 
large quantum number can be treated as a $c-$number, for its fluctuations are 
negligible and its quantum behavior is not seen. It is then possible to 
replace $a_0$ and $a_0^\dagger$ by $\sqrt{M_0}$, the square root of the total 
number of bosons in the condensate. For
completeness we review all the tedious algebraic manipulations leading
to the reduced Bogolyubov Hamiltonian\cite{bog,pines} in appendix
\ref{bogoham}:
\beq \label{redham}
\mathcal H = \frac{M_0^2}{N} U + \sum_{ {\vec k} \neq 0 } 
\tilde { \epsilon}_{\vec k}  \; 
a_{\vec k}^\dagger a_{\vec k} + \frac{ M_0 U }{ N } \sum_{ {\vec k} \neq 0 }  
( a_{\vec k}^\dagger a_{ - {\vec k} }^\dagger + a_{ - {\vec k} } a_{\vec k} )
\eneq
\noindent where $
\tilde {\epsilon}_{\vec k}  = \epsilon_{\vec k} + 4 M_0 U / N - \mu $ and 
$\varepsilon$ and $U$ factors have been subsumed in $\mu$ as the fixing of the 
average particle number will cause them to be swallowed by the chemical 
potential. 

For the ground state, which is a minimum of energy, the 
expectation value of $\mathcal H$ must satisfy
\beqs
\frac{ \partial \langle \mathcal H \rangle }{ \partial M_0 } = 0
\eneqs
\noindent If this condition is not satisfied, then by connecting a reservoir to
our system we would have a flow of particles into and out of the system until 
equilibrium is reached. As long as the low energy eigenstates do not deplete 
the condensate, which they do not for not too high $U$, they satisfy the same 
extremum condition. 
Taking the partial derivative we find
\beq \label{mu}
\mu  = \frac{ 2 M_0}{N} U
\eneq
\noindent and 
\beq \label{etilde}
\tilde { \epsilon}_{\vec k}  = \epsilon_{\vec k} + \frac{2 M_0}{N} U
\eneq

The Hamiltonian in equation (\ref{redham}) can be diagonalized by
making a Bogolyubov transformation
\beq \label{bogo}
b_{\vec k} = u_{\vec k} a_{\vec k} + v_{\vec k} a_{ - {\vec k} }^\dagger 
\qquad b_{\vec k}^\dagger = u_{\vec k} a_{\vec k}^\dagger + v_{\vec k} 
a_{ - {\vec k} }
\eneq
\noindent where we require
\beq \label{commu}
[ b_{\vec k}, b_{\vec k}^\dagger ] = u_{\vec k}^2 - v_{\vec k}^2 = 1
\eneq
\noindent in order for the transformation to be canonical. The diagonal 
Hamiltonian has the form
\beqs
\mathcal H = \sum_{\vec k} E_{\vec k} b_{\vec k}^\dagger b_{\vec k} + C
\eneqs
\noindent where $C$ is a constant. Substituting (\ref{bogo}) in the
above expression we find
\begin{multline}
\mathcal H = \sum_{\vec k} ( E_{\vec k} \; ( u_{\vec k}^2 + v_{\vec
k}^2 ) \; a_{\vec k}^\dagger a_{\vec k} + u_{\vec k} v_{\vec k}
E_{\vec k} ( a_{\vec k}^\dagger a_{ - {\vec k} }^\dagger + a_{ - {\vec
k} } a_{\vec k} ) \\
+ E_{\vec k} v_{\vec k}^2 ) + C
\end{multline}
\noindent Comparing this expression with (\ref{redham}) we identify
\beqs
E_{\vec k} \; ( u_{\vec k}^2 + v_{\vec k}^2 ) = \tilde { \epsilon}_{\vec k} 
\qquad u_{\vec k} v_{\vec k} E_{\vec k} = \frac{ M_0 }{ N } U
\eneqs
\noindent Solving these two equations together with $ u_{\vec k}^2 - 
v_{\vec k}^2 = 1 $ yields
\beq \label{uv}
u_{\vec k} = \frac{ 1 }{ \sqrt{2} } \sqrt{ \frac{ \tilde{\epsilon}_{\vec k}} 
{ E_{\vec k} } + 1 } \qquad v_{\vec k} = \frac{ 1 }{ \sqrt{2} } 
\sqrt{ \frac{ \tilde{\epsilon}_{\vec k} }{ E_{\vec k} } - 1 }
\eneq
\begin{align} \label{energy}
\nonumber E_{\vec k}^2 &= \tilde { \epsilon}_{\vec k}^2 
- \frac{4 M_0^2}{N^2} U^2 \\
E_{\vec k} &= \sqrt{\epsilon_{\vec k}}\sqrt{ \epsilon_{\vec k} 
+ 4 \frac{M_0}{N} U}
\end{align}
\noindent This is the quasiparticle excitation spectrum. 
\noindent As $ k \rightarrow 0 $
\beq
E_{\vec k} \simeq \sqrt{\epsilon_{\vec k}} \sqrt{\frac{ 4 M_0 U }{ N }} = 
p \sqrt{\frac{ 2 M_0 U }{ N m }} 
\eneq
\noindent where $ p = \sqrt{2 m \epsilon_{\vec k}} = \sqrt{2 m t} \; a k
= \hbar k$. Since as $ k \rightarrow 0 $, $ E_{\vec k} = p c_s $, i.e. low
energy excitations are sound, we identify the speed of sound to be
\beq \label{sound}
c_s = \sqrt{\frac{ 2 M_0 U }{ N m }}
\eneq
\noindent This is a finite measurable quantity that indicates the presence of 
correlations among the constituents of the material. Such correlations 
stabilize persistent currents and give rigidity to the ground state by making 
it energetically expensive to excite the fundamental bosons independently. 
Instead, the excitations are collective modes and the system superflows. The 
ground state has acquired rigidity. Artificially engineered BECs have been 
measured to have finite sound speeds\cite{bec2} and are thus superfluid.

The ground state wavefunction is worked out appendix \ref{groundstate} and 
found to be
\beq \label{gs}
\qquad \quad |\Psi_0 \rangle = \prod_{\vec k} 
\frac{ 1 }{ u_{\vec k} } e^{ -(v_{\vec k} / u_{\vec k}) a_{\vec k}^\dagger 
a_{-\vec k}^\dagger} \; (a_0^\dagger)^{M_0}|0 \rangle
\eneq
\noindent The factor $(a_0^\dagger)^{M_0}$ represents the bosons in
the condensate. In the absence of interactions, $u_{\vec k} = 1,
v_{\vec k} = 0$, so that in exciting bosons, one changes atomic boson
number by one. In the presence of interactions, states with momentum
${\vec k}$ and $-{\vec k}$ are mixed into the ground state, so when an
excitation with momentum ${\vec k}$ is created, another with momentum
$-{\vec k}$ is destroyed. The excitations are coherent superpositions
of atoms and ``holes''. This is due to the nature of the coupling
between ${\vec k}$ and $-{\vec k}$ in the Bogolyubov operators. With
interactions both $u_{\vec k}$ and $v_{\vec k}$ increase. 

The ground state contains nonseparable, nontrivial correlations between the 
bosons that make up the system, i. e., it is entangled. The correlations in 
the ground state have entangled the bosons into a coherent state for the 
lowest energy state. The entanglement is so extreme that the bosons that make 
up the system cannot be excited at long wavenumbers. Their existence at low 
energies is impossible. Only sound can be excited, i.e. the excitations are 
Bogolyubov quasiparticles which do not resemble bosons whatsoever at low 
energies. In this limit, the principles of quantum hydrodynamics become exact
because the system does not dissipate and sound excitations become exact 
eigenstates of the system. The boson fluid responds only collectively at long 
wavelengths. Aspects of the universality of superfluid physics are presented in
appendix \ref{uni}

We now proceed to calculate the self-consistency condition. Using the
ground state wave function (\ref{gs}) the self-consistency condition is
\beq
\sum_i \langle n_i \rangle = M = \langle \Psi_0 | a_{0}^\dagger 
a_{0} | \Psi_0 \rangle + \sum_{\vec k \neq0} \langle \Psi_0 | 
a_{\vec k}^\dagger a_{\vec k} | \Psi_0 \rangle
\eneq
\noindent Let $|a\rangle = a_{\vec k}|\Psi_0\rangle$, that is
\beq
|a\rangle = \prod_{ {\vec k}' \neq {\vec k} } \frac{ 1 }{ u_{ {\vec k}' } }
e^{ -\frac{v_{{\vec k}'}}{u_{{\vec k}'}} a_{{\vec k}'}^\dagger 
a_{-{\vec k}'}^\dagger } \frac{ 1 }{ u_{\vec k} }
e^{ -\frac{v_{\vec k}}{u_{\vec k}} a_{\vec k}^\dagger 
a_{-{\vec k}}^\dagger } | 0 \rangle
\eneq
\noindent The commutation relation $[a_{\vec k}, a_{\vec k}^\dagger] =
1$ implies $a_{\vec k} = \frac{\partial}{\partial a_{\bf
k}^\dagger}$. Using this result
\beq
|a\rangle = -\frac{v_{\vec k}}{u_{\vec k}} a_{-{\vec k}}^\dagger |
 \Psi_0 \rangle
\eneq
\noindent Substituting the Bogolyubov transformation $a_{ -{\vec k}
}^\dagger = u_{\vec k} b_{-{\vec k}}^\dagger - v_{\vec k} b_{\vec k}$
\beq
|a\rangle = -\frac{v_{\vec k}}{u_{\vec k}} ( u_{\vec k} b_{-{\vec k}}^\dagger 
- v_{\vec k} b_{\vec k} ) | \Psi_0 \rangle 
= - v_{\vec k} b_{-{\vec k}}^\dagger | \Psi_0 \rangle
\eneq
\noindent Thus $\langle a | a \rangle = v_{\vec k}^2$ and the
self-consistency condition becomes
\begin{align} \label{selfcons}
\nonumber M &= \sum_{\vec k} v_{\vec k}^2 + M_0 = \frac{1}{2}\sum_{\vec k}
\frac{ \tilde \epsilon_{\vec k} - E_{\vec k} }{ E_{\vec k} } + M_0 \\
&= \frac{1}{2}\sum_{\vec k} \left ( \frac{ \epsilon_{\vec k} 
+ 2 (M_0 / N)U }{ E_{\vec k} } - 1 \right ) + M_0
\end{align}
\noindent This is the ``particle conservation sum rule''.

We will now make estimates of the number of particles in the condensate as 
a function of $U/t$. We use the result (\ref{energy})
\beq \label{vk2crudo}
v_{\vec k}^2 = \frac{1}{2} \left ( \frac{\epsilon_{\vec k} + 2 (M_0 / N) U }
{ \sqrt{\epsilon_k} \sqrt{\epsilon_{\vec k} + 4 (M_0 / N) U} } - 1 \right )
\eneq
\noindent Similarly we will also estimate
\beq
u_{\vec k}^2 = \frac{1}{2} \left ( \frac{\epsilon_{\vec k} + 2 (M_0 / N) U }
{ \sqrt{\epsilon_k} \sqrt{\epsilon_{\vec k} + 4 (M_0 / N) U} } + 1 \right )
\eneq
\noindent We will concentrate on the long wavelength behavior, where
equation (\ref{epscub}) for the dispersion gives $\epsilon_{\vec k}
\simeq t k^2 a^2$. In this limit, the sum over ${\vec k}$ will be
approximated with an integral: $\sum_{\vec k} \rightarrow (N a^3/2^3
\pi^3 ) \int k^2 \; dk \; d\Omega$. Here $N$ is the number of lattice
sites.  Notice further that for $\epsilon_{\vec k} \simeq t k^2 a^2 <
2 U M_0 / N$, $v_{\vec k} \neq 0$ and $u_{\vec k} \neq 1$. On the
other hand, for $t k^2 a^2 >> 2 U M_0 / N$, $v_{\vec k} \simeq 0$ and
$u_{\vec k} \simeq 1$. This means that the system behaves like a
superfluid for long wavelengths and like a free Bose gas for short
wavelengths. Superfluidity is thus a low energy phenomenon. This
behavior can be approximated with the introduction of a cutoff, so that
above the cutoff $v_{\vec k} \simeq 0$ and $u_{\vec k}\simeq 1$ while
below the cutoff they are given by the integral through the region $0
\leq k \leq k_c $. The cutoff is given by
\beq
t k_c^2 a^2 = 2 \frac{U M_0}{N} \qquad ; 
\qquad k_c = \frac{\sqrt{2}}{a} \sqrt{ \frac{U M_0}{t N} }
\eneq
\noindent For $k < k_c$ the system responds like a superfluid while
otherwise it behaves like a dissipative free Bose gas. The important
detail is that as long as the cutoff wavevector is smaller than the
absolute cutoff from the lattice, $\pi/ a$, so that $\frac{2 U M_0}{t
N \pi^2} \leq 1$, it will define the cutoff for all momentum integrals
is $k_c$. The system generated a cutoff scale from the physics that
lead to superfluidity. This scale is related to the all important
coherence length of the superfluid about which we will have a lot to
say later. Going back to (\ref{vk2crudo}) and writing $\epsilon_{{\vec
k}_c}$ for $2 U M_0 / N $
\begin{align}
\nonumber v_{\vec k}^2 &= \frac{1}{2} \left ( \frac{ \epsilon_{\vec k} 
+ \epsilon_{{\vec k}_c}}
{ \sqrt{\epsilon_k} \sqrt{\epsilon_{\vec k} 
+ 2\epsilon_{{\vec k}_c} } } - 1 \right )  \\
\label{vk2} & \simeq \frac{1}{2} \left ( \frac{1}{\sqrt{2}} 
\sqrt{ \frac{ \epsilon_{{\vec k}_c} }
{ \epsilon_{\vec k} } } - 1 \right )
\end{align}
\noindent and
\beq \label{uk2}
u_{\vec k}^2 \simeq \frac{1}{2} \left ( \frac{1}{\sqrt{2}} 
\sqrt{ \frac{ \epsilon_{{\vec k}_c} }
{ \epsilon_{\vec k} } } + 1 \right )
\eneq
\noindent In terms of the cutoff 
\beq
v_{\vec k}^2 \simeq \frac{1}{2} \left ( \frac{1}{\sqrt{2}} 
\frac{ k_c }{ k } \right ) \simeq u_{\vec k}^2 
\eneq
\noindent Thus
\beq \label{intv^2}
\sum_{\vec k} v_{\vec k}^2 \simeq \frac{N a^3 k_c}{16 \sqrt{2} \pi^3} 
\int_0^{k_c} \frac{1}{k} \; d^3{\vec k} = \frac{1}{8 \sqrt{2} \pi^2}N a^3 k_c^3
\eneq
\noindent Substituting this into (\ref{selfcons}) gives for the
self-consistency condition
\beq
M = \frac{1}{8 \sqrt{2} \pi^2}N a^3 k_c^3 + M_0
\eneq
\noindent This provides an expression for $U / t$ in terms of $M_0 / M$:
\begin{align}
\nonumber 1 &= \frac{1}{8\sqrt{2}\pi^2}a^3k_c^3 \frac{N}{M} + \frac{M_0}{M} \\
\nonumber 1 &= \frac{1}{8 \sqrt{2} \pi^2} \left( 
\frac{2 M_0 U}{N t} \right)^{3/2} \frac{N}{M} + \frac{M_0}{M} \\
\frac{U}{t} &= ( 8 \sqrt{2} \pi^2 )^{2/3} 
\frac{N}{2 M_0} \left ( \frac{M}{N} \right )^{2/3}  
\left ( 1 - \frac{M_0}{M} \right )^{2/3}
\end{align}

We now proceed to calculate the expected value of the Bose Hubbard
Hamiltonian in the superfluid ground state in order to compare with
the energy of the other phases and see which one is energetically more
favorable. One of the main concerns of the present work is the Mott
transition. So in particular we will compare this ground state energy
of the superfluid phase with a certain Mott phase not previously
studied in the literature\cite{fisher}. The full Bose Hubbard
Hamiltonian is given in equation (\ref{fullham}). The expectation
value of the energy in the superfluid ground state is then
\begin{multline} \label{expham}
\langle \mathcal H \rangle = - ( \varepsilon + U ) M + \sum_{\vec k} 
\epsilon_{\vec k} v_{\vec k}^2 \\
+ \frac{ U }{ N } \sum_{ {\vec k}_1, {\vec k}_2 }
\sum_{\vec q} \langle a_{ {\vec k}_1 + {\vec q} }^\dagger a_{ {\vec k}_1 } 
a_{ {\vec k}_2 - {\vec q} }^\dagger a_{ {\vec k}_2 } \rangle 
- 6 t M
\end{multline}
\noindent Notice that $a_{ {\vec k}_1 } a_{ {\vec k}_2 - {\vec q} } =
\delta_{ {\vec k}_1, {\vec k}_2 - {\vec q} } + a_{ {\vec k}_2 - {\vec
q} }^\dagger a_{ {\vec k}_1 }$. Using this result we obtain for the
interaction term
\begin{align} \label{expint}
\nonumber \langle \mathcal{H}_I \rangle &= \frac{ U }{ N } 
\sum_{ {\vec k}_2, {\vec q} } a_{ {\vec k}_2 }^\dagger a_{ {\vec k}_2 }
+ \frac{ U }{ N } \sum_{ {\vec k}_1, {\vec k}_2 } \sum_{\vec q} \langle 
a_{ {\vec k}_1 + {\vec q} }^\dagger  a_{ {\vec k}_2 - {\vec q} }^\dagger
a_{ {\vec k}_1 } a_{ {\vec k}_2 } \rangle \\
&= U M + \frac{U}{N} \sum_{ {\vec k}_1, {\vec k}_2 } 
\sum_{\vec q} \langle a_{ {\vec k}_1 + {\vec q} }^\dagger  
a_{ {\vec k}_2 - {\vec q} }^\dagger
a_{ {\vec k}_1 } a_{ {\vec k}_2 } \rangle
\end{align}
\noindent On the other hand we have from (\ref{aasfgs})
\begin{align}
\nonumber a_{ {\vec k}_1 } a_{ {\vec k}_2 } | \Psi_0 \rangle 
&= \frac{ - v_{ {\vec k}_2} }{ u_{ {\vec k}_2 } }
( 1 + \frac{v_{{\vec k}_1}^2 u_{{\vec k}_2}}{u_{{\vec k}_1}} ) 
\delta_{ {\vec k}_1 , - {\vec k}_2 } | \Psi_0 \rangle \\
\nonumber & \qquad \qquad + v_{ {\vec k}_1} v_{ {\vec k}_2 } 
b_{ - {\vec k}_1 }^\dagger b_{ - {\vec k}_2 }^\dagger | \Psi_0 \rangle
\end{align}
\noindent Similarly
\begin{align}
\nonumber \langle \Psi_0 | a_{ {\vec k}_1 + {\vec q} }^\dagger 
a_{ {\vec k}_2 - {\vec q} }^\dagger 
&= \frac{ - v_{ {\vec k}_1 + {\vec q} } }{ u_{ {\vec k}_1  + {\vec q} } } 
( 1 + \frac{v_{{\vec k}_2 - {\vec q}}^2 u_{{\vec k}_1 + {\vec q}}}
{u_{{\vec k}_2 - {\vec q}}} )
\langle \Psi_0 | \; \delta_{ {\vec k}_1, - {\vec k}_2 } \\
\nonumber & + v_{ {\vec k}_2 - {\vec q} } v_{ {\vec k}_1 + {\vec q} }  
\langle \Psi_0 | \; b_{ - {\vec k}_1 - {\vec q} } 
b_{ - {\vec k}_2 + {\vec q} }  
\end{align}
\noindent Joining the two results we finally get
\begin{widetext}
\begin{align}
\nonumber \langle a_{ {\vec k}_1 + {\vec q} }^\dagger 
a_{ {\vec k}_2 - {\vec q} }^\dagger a_{ {\vec k}_1 } a_{ {\vec k}_2 } \rangle 
&= \frac{ v_{ {\vec k}_2} }{ u_{ {\vec k}_2 } } 
\frac{ v_{ {\vec k}_1 + {\vec q} } }{ u_{ {\vec k}_1  + {\vec q} } }
( 1 + \frac{v_{{\vec k}_1}^2 u_{{\vec k}_2}}{u_{{\vec k}_1}} ) 
( 1 + \frac{v_{{\vec k}_2 - {\vec q}}^2 u_{{\vec k}_1 + {\vec q}}}
{u_{{\vec k}_2 - {\vec q}}} )
\delta_{ {\vec k}_1, - {\vec k}_2 } 
+ v_{ {\vec k}_1 } v_{ {\vec k}_2 } v_{ {\vec k}_2 - {\vec q} }
v_{ {\vec k}_1 + {\vec q} }
\langle b_{ - {\vec k}_1 - {\vec q} } b_{ - {\vec k}_2 + {\vec q} } 
b_{ - {\vec k}_1 }^\dagger b_{ - {\vec k}_2 }^\dagger \rangle \\
\nonumber &= \frac{ v_{ {\vec k}_2} }{ u_{ {\vec k}_2 } } 
\frac{ v_{ {\vec k}_1 + {\vec q} } }{ u_{ {\vec k}_1  + {\vec q} } }
( 1 + \frac{v_{{\vec k}_1}^2 u_{{\vec k}_2}}{u_{{\vec k}_1}} ) 
( 1 + \frac{v_{{\vec k}_2 - {\vec q}}^2 u_{{\vec k}_1 + {\vec q}}}
{u_{{\vec k}_2 - {\vec q}}} )
\delta_{ {\vec k}_1, - {\vec k}_2 } \\
& \qquad \qquad \qquad + v_{ {\vec k}_1 } v_{ {\vec k}_2 } v_{ {\vec
k}_2 - {\vec q} } v_{ {\vec k}_1 + {\vec q} } ( \delta_{ - {\vec k}_2
+ {\vec q} , - {\vec k}_1 } \delta_{ - {\vec k}_1 - {\vec q} , - {\vec
k}_2 } + \delta_{ - {\vec k}_2 + {\vec q} , - {\vec k}_2 } \delta_{ -
{ {\vec k}_1 } - {\vec q} , - {\vec k}_1 } )
\end{align}
\end{widetext}
\noindent Thus performing the sums
\begin{align}
\nonumber \sum_{{\vec k}_1, {\vec k}_2} \sum_{\vec q} 
\langle a_{ {\vec k}_1 + {\vec q} }^\dagger 
& a_{ {\vec k}_2 - {\vec q} }^\dagger 
a_{ {\vec k}_1 } a_{ {\vec k}_2 } \rangle 
\qquad \qquad \qquad \qquad \qquad \qquad \\
\label{sums} \nonumber &= \sum_{{\vec k}, {\vec k}_1} 
\frac{ v_{\vec k} }{ u_{\vec k} } 
\frac{ v_{{\vec k}_1} }{ u_{{\vec k}_1} } 
( 1 + v_{\vec k}^2 ) ( 1 + v_{{\vec k}_1}^2 ) \\
& \qquad \qquad + 2 \sum_{{\vec k}, {\vec k}'} 
v_{\vec k}^2 v_{ {\vec k}' }^2
\end{align}
\noindent where ${\vec k}_1 \equiv {\vec k} - {\vec q}$. From (\ref{vk2}) and 
(\ref{uk2}) $u_{\vec k}^2 / v_{\vec k}^2= 1$. Hence
\beq \label{intv/u}
\sum_{\vec k} \frac{ v_{\vec k} }{ u_{\vec k} } 
\simeq \frac{N a^3}{2^3  \pi^3} \int_0^{k_c} 
\; d^3{\vec k} = \frac{ 1 }{ 6 \pi^2 } N a^3 k_c^3
\eneq
\noindent Usage of (\ref{intv^2}) and (\ref{intv/u})
into (\ref{sums}) gives
\begin{align} \label{fsum}
\nonumber \sum_{{\vec k}_1, {\vec k}_2} \sum_{\vec q} 
\langle & a_{ {\vec k}_1 + {\vec q} }^\dagger 
a_{ {\vec k}_2 - {\vec q} }^\dagger 
a_{ {\vec k}_1 } a_{ {\vec k}_2 } \rangle \\
& \simeq \left ( \frac{1}{9} + \frac{1}{6 \sqrt{2}} +\frac{1}{32}
\right ) \frac{N^2 a^6 k_c^6}{4 \pi^4}
\end{align}
\noindent As for the kinetic energy, with $\epsilon_{\vec k} \simeq t
k^2 a^2$ and (\ref{vk2})
\begin{align} \label{kinen}
\nonumber \sum_{\vec k} \epsilon_{\vec k} v_{\vec k}^2 
&\simeq \frac{N a^3}{16\sqrt{2} \pi^3}  t a^2 k_c 
\int_0^{k_c} k \; d^3{\vec k} \\
&= \frac{1}{16\sqrt{2} \pi^2} N t a^5 k_c^5
\end{align}
\noindent Substituting (\ref{fsum}), (\ref{expint}) and (\ref{kinen}) into
(\ref{expham}) gives
\begin{align}
\nonumber \langle \mathcal H \rangle \simeq &- \varepsilon M 
+  \frac{1}{16\sqrt{2} \pi^2} N t a^5 k_c^5\\
\nonumber & + N U\left ( \frac{1}{9} + \frac{1}{6 \sqrt{2}} +\frac{1}{32} 
\right )
 \frac{ a^6 k_c^6}{4 \pi^4} \\
&- 6 t M \nonumber \\
\nonumber =&- \varepsilon M 
+  \frac{1}{16\sqrt{2} \pi^2} N t \left( \frac{2 M_0 U}{N t}\right)^{5/2}\\
\nonumber & + \left ( \frac{1}{9} + \frac{1}{6 \sqrt{2}} +\frac{1}{32} \right )
 \frac{N U}{4 \pi^4} \left( \frac{2 M_0 U}{N t} \right)^3\\
&- 6 t M \label{esf}
\end{align}
\noindent If the number of bosons is an integer multiple of the number
of lattice sites, the system is said to be commensurate. On the other
hand, if the number of bosons is not an integer multiple of the number
of lattice sites, the system is said to be incommensurate. Notice that
the above result is general and thus applies for both incommensurate
and commensurate systems. 

In the limit of small $U / t$, $M_0 \simeq M$ and the second and third
terms are small. The energetically expensive term, $U N$ is
small. There is an energy gain in becoming superfluid. As $U / t$
increases $\frac{U M_0}{t N}$ goes to a constant as we will see in a
succeeding section, and the second and third become of the order of
$N$. If the second and third terms cancel the kinetic energy term
$-6tM$, the system is not a quantum fluid anymore. It will thus have a
transition into an insulating phase, a Mott phase. The
commensurability does not play a role in the depletion of the
condensate, contrary to what was found in the earlier work
\cite{fisher}. The earlier work is not incorrect. In that work, the
Mott transition happens when the superfluid within each well dephases,
preventing Josephson tunneling and leading to a Mott-phase. That
transition is continuous and only possible for commensurate
values. For this to happen there needs to exist a superfluid within
each well, which requires enough bosons per well (broken symmetries
can only happen for a macroscopic number of particles) and the
coherence length to be longer than the well. So in certain physical
situations there is a Mott transition of a different type,
discontinuous and incommensurate, since the superfluid does not survive
within each well, and individual bosons are localized in each well. We
will elucidate these physics in following sections.

\section{Order Parameter in the Superfluid}

In the present section we go on with the study of the universal
properties of the superfluid. We concentrate in what is the superfluid
order parameter. This section also provides a connection with the new
kind of Mott transition we have uncovered in our work. In particular,
since we will find that the superfluid order parameter increases as we
increase the onsite repulsion $U$, i.e. as we approach a Mott
transition, a transition in which the superfluid order parameter is
destroyed must be discontinuous. The unique experimental signature of
this is a discontinuous jump in any measure of the superfluid response.

We have seen that the well depth $\varepsilon$ is subsumed into the
chemical potential upon minimization of the ground state energy with
respect to the number of particles in the condensate. The physical
reason for this is that whether the lattice is deep or shallow is
irrelevant as long as the system is superfluid; it superflows through
it. Nonetheless, the well depth is relevant to the Mott transitions
and phases. Also, the kinetic term in (\ref{hami}) depends on
$\varepsilon$. This is so because decreasing the depth of the
potential well, i.e. increasing $\varepsilon$, will obviously decrease
$t$ as the barrier through which tunneling must occur is higher. The only
vestige of $\varepsilon$ remaining in the superfluid is through
$t(\varepsilon)$. We also notice that $U/t \rightarrow \infty$ with
decreasing $t$. This however does not kill the order parameter of the
BEC as long as the condensate is not depleted or Josephson tunneling
suppressed. If the condensate is depleted there will be a Mott
transition in the universality class of the one uncovered in the
present work. If Josephson tunneling is suppressed without destruction
of the order parameter, there will be a Mott transition in the
universality class discovered in the previous class \cite{fisher}.

It is commonly said in the literature that the existence of a
condensate is essential for superfluidity. The macroscopic occupation
of the lowest momentum state, i.e. the presence of the condensate,
makes the boson system into a quantum fluid as long as the separation
between energy levels is unmeasurable. The existence of the condensate
does not imply the existence of a superfluid. Repulsive interactions
imply the existence of superfluidity as long as there is a condensate,
i.e. the system is fluid. We have seen that Bogolyubov correlations
are the consequence of the interactions which leads to the necessary
rigidity for superfluidity. Therefore, the correct order parameter
that characterizes a superfluid is $\langle a_{\vec k} a_{ - {\vec k}
} \rangle$, which is non zero only when there are Bogolyubov
correlations.

Now that we have recognized the correct order parameter, the direct
relationship with BCS superconductivity is obvious. {\it The
superfluid state is universal.} Be it because of a BEC that acquires
rigidity due to the presence of repulsive interactions, or be it
because of a Fermi gas acquiring rigidity due to attractive
interactions, once the system orders into a superfluid, the
macroscopic end result in both systems is the same, independent of
microscopics, modulo charging effects. There is an energy gain by
ordering which stabilizes superflow and in both cases it is caused by
the action of interactions. Both systems exhibit a Meissner effect:
the BCS ground state does not respond to small enough magnetic fields
while the {\it uncharged} superfluid BEC system does not respond to
slow enough rotations.
\begin{figure}[ht]  
\centering
\resizebox{8cm}{!}{%
   \includegraphics*{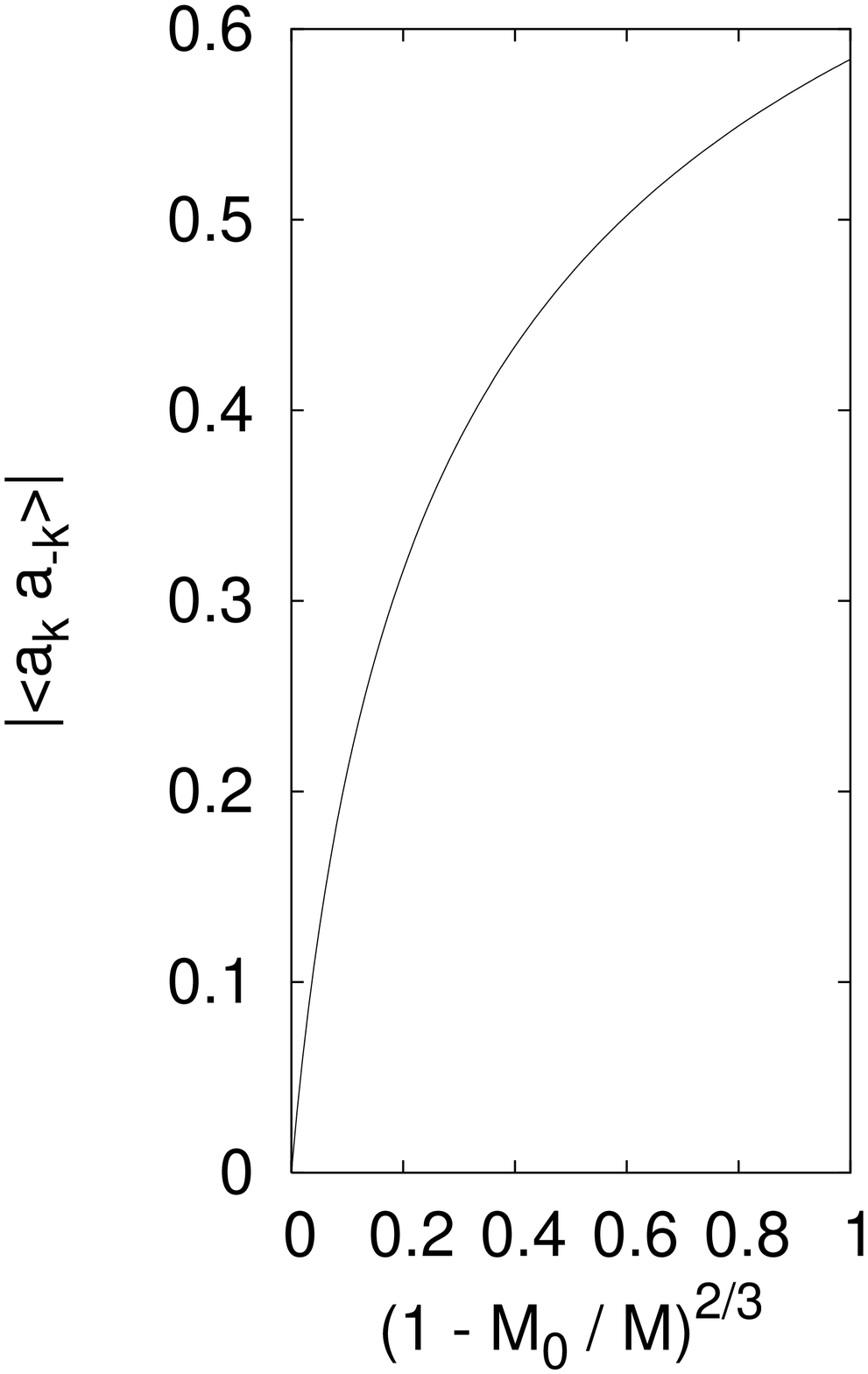}\includegraphics*{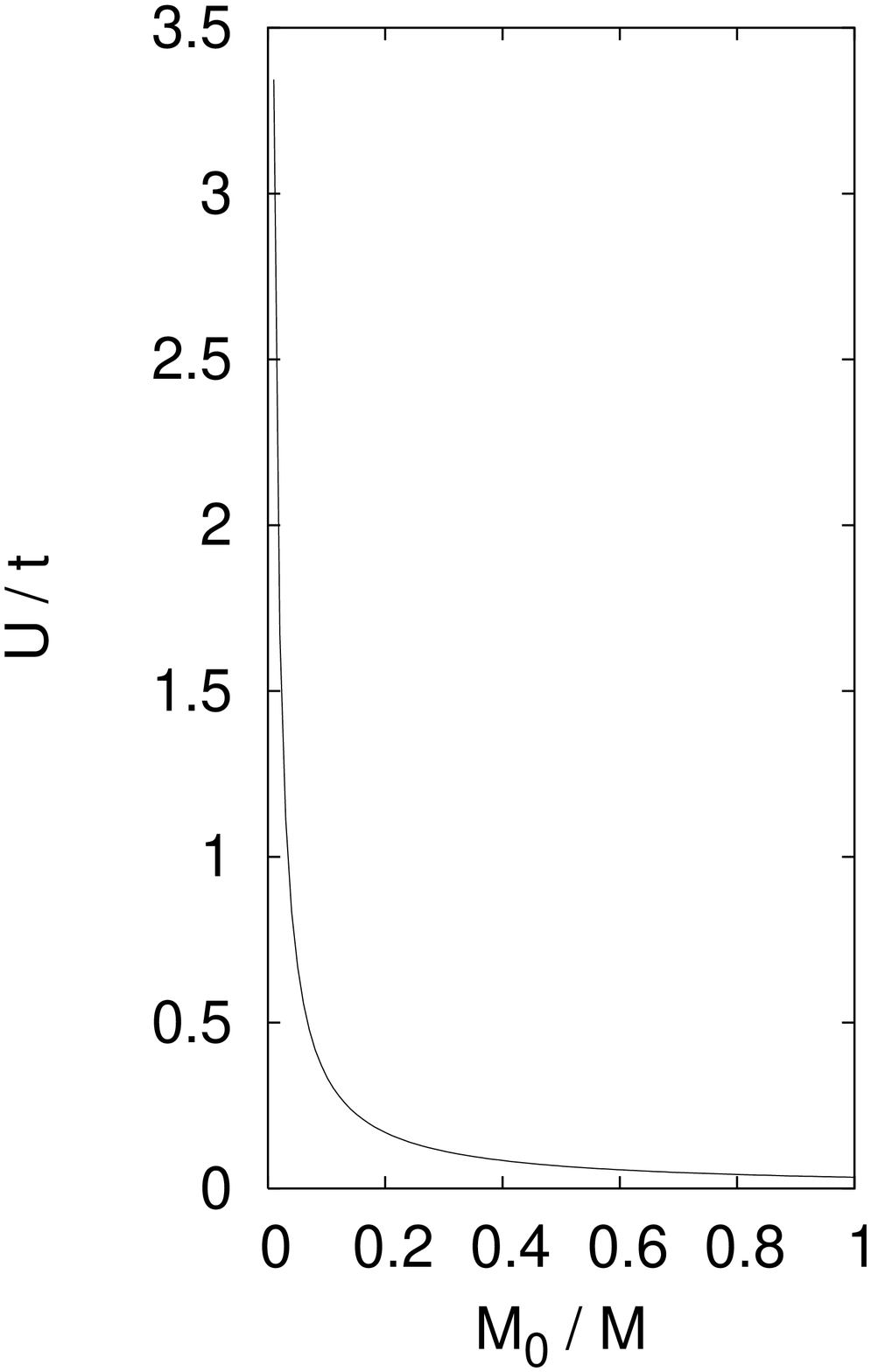}}    
\caption{\label{odplot} (a) Dependence of the order parameter on the
fraction of bosons in the condensate, (b) $U / t$ vs. $M_0 / M$}
\end{figure}

The normalized Bogolyubov superfluid ground state wavefunction is
\beq 
|\Psi_0 \rangle = \prod_{\vec k} \frac{ 1 }{ u_{\vec k} } e^{
  -(v_{\vec k} / u_{\vec k}) a_{\vec k}^\dagger a_{-\vec k}^\dagger}
\; (a_0^\dagger)^{M_0} |0 \rangle
\eneq
\noindent Thus the expectation value of the superfluid order parameter
is given by
\beq
\langle a_{\vec k} a_{ - {\vec k} } \rangle = \frac{ 1 }{ u_{\vec k}^2 } 
\langle 0 | a_0^{M_0} \; e^{ -(v_{\vec k} / u_{\vec k}) a_{ - {\vec k} } 
a_{\vec k} } a_{-\vec k} a_{\vec k} e^{ -(v_{\vec k} / u_{\vec k}) 
a_{\vec k}^\dagger a_{ - {\vec k} }^\dagger} \; (a_0^\dagger)^{M_0} |0 \rangle
\eneq
\noindent where
\beq
a_{-\vec k} a_{\vec k} e^{ -(v_{\vec k} / u_{\vec k}) 
a_{\vec k}^\dagger a_{ - {\vec k} }^\dagger} = \sum_n 
\frac{ a_{-\vec k} a_{\vec k} }{ n! } 
( \frac{ - v_{\vec k} }{ u_{\vec k} } )^n ( a_{\vec k}^\dagger 
a_{-\vec k}^\dagger )^n
\eneq
\noindent Since $[a_{-\vec k} a_{\vec k}, (a_{\vec k}^\dagger)^n 
(a_{-\vec k}^\dagger)^n] = n ( a_{\vec k}^\dagger 
a_{-\vec k}^\dagger )^{ n - 1 }$ we find
\begin{align}
\nonumber \langle a_{\vec k} a_{ - {\vec k} } \rangle &= \frac{ 1 }{
u_{\vec k}^2 } \langle 0 | a_0^{M_0} \; e^{ -(v_{\vec k} / u_{\vec k})
a_{ - {\vec k} } a_{ - {\vec k} } } \\
\nonumber &\times \sum_n ( \frac{ - v_{\vec k} }{ u_{\vec k} } )^n
\frac{ ( a_{\vec k}^\dagger a_{-\vec k}^\dagger )^{ n - 1 } } { (n -
1)! } (a_0^\dagger)^{M_0} |0 \rangle \\
\nonumber &= \frac{ - v_{\vec k} }{ u_{\vec k} } \left( \langle 0 | \;
a_0^{M_0} e^{ -(v_{\vec k} / u_{\vec k}) a_{ - {\vec k} } a_{\vec k} }
e^{ -(v_{\vec k} / u_{\vec k}) a_{\vec k}^\dagger a_{ - {\vec k}
}^\dagger } \; (a_0^\dagger)^{M_0} | 0 \rangle \right) \\
&= \frac{ - v_{\vec k} }{ u_{\vec k} } \\
\nonumber &= - \sqrt{ \frac{ (\tilde \epsilon_{\vec k} / E_{\vec k}) - 1 }
{\tilde \epsilon_{\vec k} / E_{\vec k}) + 1} } \\
\langle a_{\vec k} a_{ - {\vec k} } \rangle &= - \sqrt{ \frac{
\epsilon_{\vec k} + \frac{2 M_0 U}{N} - E_{\vec k} }{ \epsilon_{\vec
k} + \frac{2 M_0 U}{N} + E_{\vec k} } }
\end{align}
\noindent since $E_{\vec k} = \sqrt{\epsilon_{\vec k}} 
\sqrt{ \epsilon_{\vec k} + \frac{4 M_0 U}{N} }$ we find
\beq
| \langle a_{\vec k} a_{ - {\vec k} } \rangle | = \sqrt{ \frac{
    \epsilon_{\vec k} + \frac{2 M_0 U}{N} - \sqrt{\epsilon_{\vec k}}
    \sqrt{ \epsilon_{\vec k} + \frac{4 M_0 U}{N} } }{ \epsilon_{\vec
    k} + \frac{2 M_0 U}{N} + \sqrt{\epsilon_{\vec k}} \sqrt{
    \epsilon_{\vec k} + \frac{4 M_0 U}{N} } } }
\eneq
\noindent or with $\epsilon_{\vec k} \simeq t k^2 a^2$ and
$\frac{2M_0U}{Nt} = \left(\frac{8 \sqrt{2} \pi^2 M}{N}\right)^{2/3} (
1 - M_0/M )^{2/3}$
\begin{widetext}
\beq
| \langle a_{\vec k} a_{ - {\vec k} } \rangle | = \sqrt{ \frac{ k^2
    a^2 + \left(\frac{8 \sqrt{2} \pi^2 M}{N}\right)^{2/3} ( 1 - M_0/M
    )^{2/3} - k a \sqrt{k^2 a^2 + 2 \left(\frac{8 \sqrt{2} \pi^2
    M}{N}\right)^{2/3} ( 1 - M_0/M )^{2/3} } }{ k^2 a^2 +
    \left(\frac{8 \sqrt{2} \pi^2 M}{N}\right)^{2/3} ( 1 - M_0/M
    )^{2/3} + k a \sqrt{ k^2 a^2 + 2 \left(\frac{8 \sqrt{2} \pi^2
    M}{N}\right)^{2/3} ( 1 - M_0/M )^{2/3} } } }
\eneq
\end{widetext}

As the potential depth is increased, the tunneling amplitude, and
hence $t$ in the Hamiltonian, decreases. This does not degrade the
order parameter. We also expect this to stabilize energetically the
Mott insulating phase in order to minimize local number
fluctuations. We thus expect a discontinuous transition into a Mott
insulating phase, where the order parameter experiences a sudden jump
to 0 and the individual bosons, not the pairs, get localized into the
potential wells. 

A superfluid has a density of Bogolyubov pairs. This means that even
if only a few bosons are paired, the whole system superflows because
of the rigidity acquired by pairing. We plot the dependence of the
order parameter on the number of bosons in the condensate and on $U/t$
on figure \ref{odplot}.

\section{Mott Insulator and Mean Field Solutions}

The superfluid-Mott transition is between a superfluid and a quantum
solid. As we have mentioned before in the present article and in
earlier work \cite{us} there are two types of Mott phases and Mott
transitions. One of them exists when the superfluid survives within
each well, both Josephson tunneling is suppressed within each well
leading to insulating behavior. The other which we have uncovered and
it is partly the purpose of this work to explain occurs when the
onsite repulsion destroys the superfluid order parameter and leads to
boson localization in the wells. In systems without disorder,
fluid-solid transitions are usually discontinuous as the order
parameters are too different \cite{landau2}. In the superfluid system
that is certainly the case.  Since our Mott insulating phase is
characterized by a well defined number of bosons per site, one could
take the density as its order parameter. As it is very hard to have
continuous transitions between a solid and a liquid, the continuous
transition studied in the earlier work \cite{fisher} only happens for
a commensurate number of bosons per site. This is true, but whether a
discontinuous transition can also occur at incommensurate values is
not so clear. In a following section we will study tunneling between
wells in order to elucidate the difference of the two Mott phases and
transitions and speculate as to whether there might be an
incommensurate transition in the case of suppression of Josephson
tunneling.

The Hamiltonian (\ref{hami}) with strong enough repulsion has a Mott
insulating ground state. We rewrite equation (\ref{hami}) as
\beq \label{hami2}
\mathcal H = -t \sum_{ \delta, i } a_{i + \delta}^\dagger a_i 
- ( \varepsilon + U ) \sum_i n_i + U \sum_i n_i^2
\eneq

In the case where $M < N$ there is no site with double
occupancy. States with double occupancy are excited states. This is
not true if $M > N$. We treat the two cases separately. The strong
repulsion ground state, or Mott insulator, wavefunction corresponds to
a rigid lattice with certain number of boson per lattice site,
depending on the incommensuration. For $M < N$ it is
\beq \label{M-}
|\psi_0 \rangle = \prod_{j = 1}^M a_{i_j}^\dagger | 0 \rangle
\eneq
\noindent The double subscript notation helps differentiate the sites
with bosons present (which are filled as $j$ runs from $1$ to $M$)
from the empty sites (the ones left after all the values of $j$ have
been covered). The ground state wavefunction for $p N < M < (p + 1) N$ is
\beq \label{M+}
|\psi_0 \rangle = \prod_{j = 1}^{M-pN} a_{i_j}^\dagger 
\prod_{l = 1}^N (a_l^\dagger)^p | 0 \rangle
\eneq
\noindent With the Hamiltonian given by (\ref{hami2}) the energy of
the Mott state is found to be
\begin{multline}
\langle \psi_0 | \mathcal H | \psi_0 \rangle = -t \sum_{ i, \delta } 
\langle \psi_0 | a_{i + \delta}^\dagger a_i | \psi_0 \rangle \\
- (\varepsilon + U) \sum_i \langle \psi_0 | n_i | \psi_0 \rangle 
+ U \sum_i \langle \psi_0 | n_i^2 | \psi_0 \rangle 
\end{multline}
\noindent The first term corresponds to the kinetic energy. It
destroys a boson in site $i$ and creates it in its nearest neighbor
site $i + \delta$. Thus the overlap of the state resulting from this
operation in the ground state with the original ground state gives
0. For $M < N$ usage of the ground state wavefunction (\ref{M-}) in
the above equation gives
\beq \label{E-}
\langle \psi_0 | \mathcal H | \psi_0 \rangle = 
- (\varepsilon + U) M + U M = - \varepsilon M
\eneq
\noindent On the other hand, for $p N < M < (p + 1) N$, usage of the
ground state wavefunction (\ref{M+}) gives
\begin{multline}
\nonumber \langle \psi_0 | \mathcal H | \psi_0 \rangle 
= - (\varepsilon + U) M 
+ U (M - p N) (p + 1)^2 \\
+ U [N - (M - p N)] p^2 \\
\end{multline}
\beq \label{E+}
\langle \psi_0 | \mathcal H | \psi_0 \rangle 
= - \varepsilon M + 2 p U M - (p + 1) p N U 
\eneq
\noindent This energy is to be compared with the ground state energy
of the superfluid (\ref{esf}). For $U/t$ small, we see that the
interaction terms in the superfluid are almost irrelevant when
compared with those of the Mott phase so the superfluid state is the
stable one. We also notice that for not too large an $M$ the
interaction terms in the superfluid grow a lot faster than the
interaction terms in the Mott insulating phase. The physical reason
for this behavior is that in the fluid all particles overlap with
each other as there is a probability for each of them to be at each
lattice site. Therefore for some critical $U/t$ the Mott insulating
phase will become stable causing the system to undergo a quantum phase
transition. Whether $M$ is commensurate or incommensurate with the
lattice is irrelevant to the energetics: The state with {\it all}
bosons localized is energetically favorable to the state with {\it
all} bosons superfluid. Whether the instability of the superfluid is
toward a phase with all bosons localized, or into a state with
coexistence of a commensurate Mott insulator with the leftover bosons
superfluid is not straightforward to answer variationally. It can be
answered experimentally as in the case of coexistence the superfluid
response will have a discontinuous jump to a smaller {\it nonzero}
value, while in the case of all bosons localized the superfluid
response will have a discontinuous jump to 0.

Let us take a look at the expectation value of the order parameter in
the Mott phase. It is given by
\beq
\langle a_{\vec k} a_{ - {\vec k} } \rangle = \frac{ 1 }{ N } \sum_{<ij>} 
e^{ i {\vec k} \cdot ( {\bf r}_i - {\bf r}_j ) } \langle \psi_0 | a_i a_j | 
\psi_0 \rangle
\eneq
\noindent where
\beq
\langle 0 | \prod_l a_l^{q_l} a_i a_j 
\prod_l (a_l^\dagger)^{q_l} | 0 \rangle 
= \prod_{l \neq i, j} \langle 0 | F_{ij} a_l^{q_l} 
(a_l^\dagger)^{q_l} | 0 \rangle 
\eneq
\noindent In the expression above, $q_l$ indicates the number of
bosons in site $l$ and $F_{ij}$ is a function of bosonic operators on
sites $i$ and $j$. 
\begin{align}
\nonumber \prod_{l \neq i, j} \langle 0 | F_{ij} a_l^{q_l} 
(a_l^\dagger)^{q_l} | 0 \rangle 
&= \prod_{l \neq i, j} \langle 0 | F_{ij} a_l^{q_l - 1} 
a_l (a_l^\dagger)^{q_l} | 0 \rangle \\
\nonumber &= \prod_{l \neq i, j} \langle 0 | F_{ij} a_l^{q_l - 1} 
q_l (a_l^\dagger)^{q_l - 1} | 0 \rangle \\
\nonumber & + \prod_{l \neq i, j} \langle 0 | F_{ij} a_l^{q_l - 1} 
(a_l^\dagger)^{q_l} a_l | 0 \rangle \\
&= \prod_{l \neq i, j} \langle 0 | F_{ij} a_l^{q_l - 1} 
q_l (a_l^\dagger)^{q_l - 1} | 0 \rangle 
\end{align}
\noindent where we have used $a_l = \frac{\partial}{\partial
a_l^\dagger}$. Continuing the process we find
\begin{align}
\nonumber \prod_{l \neq i, j} \langle 0 | F_{ij} & a_l^{q_l - 1} 
q_l (a_l^\dagger)^{q_l - 1} | 0 \rangle 
= \prod_{l \neq i, j} q_l ! \langle 0 | F_{ij} | 0 \rangle \\
&= \prod_{l \neq i, j} q_l ! \langle 0 | a_i^{q_i} a_j^{q_j} 
a_i a_j (a_i^\dagger)^{q_i} (a_j^\dagger)^{q_j} | 0 \rangle 
\end{align}
\noindent Since there is a non matching number of operators acting on
the ground state this yields 0. The order parameter is 0 in the Mott
phase
\beq
\langle a_{\vec k} a_{ - {\vec k} } \rangle = 0
\eneq
\noindent Actually this is obvious because the Mott state has a well
defined number of bosons per site and number phase uncertainty makes
the off-diagonal order zero. Since as $U$ increases, the order
parameter increases in the superfluid phase and the order parameter is
zero in the Mott phase, the only way to go into the Mott insulating
phase is to have a discontinuous change in the superfluid order
parameter. The Mott transition is discontinuous, i.e. first order,
with the superfluid order parameter, and thus the superfluid response,
experiencing a sudden jump to 0.

\section{Incommensurate Mott Phase}

The problem of the phase transition in a BEC has been treated in the
not distant past by M.P. Fisher, {\it et. al.} \cite{fisher}. Their
treatment relies in having a Bose condensate within each well and the order 
parameter $\langle a_{\vec k} a_{ - {\bf
k} } \rangle$ being fixed while the phase fluctuates more and more as
the superfluid density stiffens up due to the increase in $U / t$. It
is interesting to note that despite the increase of the order
parameter with decreasing $t$, the speed of sound will be
collapsing. This is the loss of rigidity due to phase fluctuations
that we just mentioned, which signals a transition into a Mott phase
{\it different} than the Mott insulating phase studied in the last
section. The Mott phase of Fisher, {\it et. al.} will be a solid with
no long range coherence but superfluid coherence within each well.
The transition out of the superfluid could be a continuous transition
as the phases at different lattice sites lose coherence and
corresponds to the transition studied by Fisher {\it et. al.}. It
might be possible for this transition to be discontinuous for
incommensurate number of bosons per lattice sites. We will try to
study this matter in the near future.

While this transition can certainly occur, the transition which takes
place in most experiments in optical lattices corresponds to
individual boson localization into a Mott insulating phase as studied
in the previous section. The reason for this is that most experiments
have low superfluid densities and equivalently, small number of bosons
per site \cite{ib,ari}. Since one needs a macroscopic number of bosons
to have a superfluid and one needs coherence lengths shorter than the
well size in order for superfluidity to survive within each well. The
experimental conditions of most lattice experiments cannot support a
transition as studied in the work of Fisher {\it et. al.} in which the
superfluid order parameter does not die.

\begin{figure}[ht]
\centering \resizebox{7cm}{!}{%
  \includegraphics*{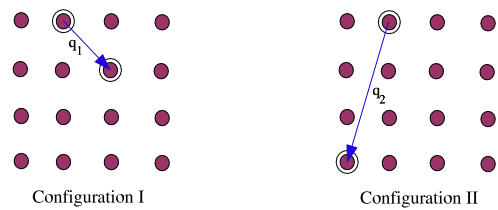}}
\caption{\label{incom} Two possible configurations for extra bosons in
an incommensurate lattice.}
\end{figure}

In our transition, when the individual bosons localize, the Bogolyubov
boson pair correlations are broken apart within the lattice and within
each well. In a commensurate system, if the depth of the potential well
is enough to overcome the on site repulsion $U$, the system will
localize and each site will be filled with the same number of
bosons. The maximum lattice vector will be $\frac{2\pi}{a}$. For an
incommensurate system the situation is somewhat different. Suppose
there are $N$ lattice sites, but $N + \frac{2}{3}$ bosons. Then as the
depth of the potential well is increased $U$ increases and $N$ bosons
will be localized one per site, forming an underlying lattice. There
are several possibilities for the remaining $\frac{2}{3}$ bosons to
go. Two of them are shown in the Figure \ref{incom}.

If only onsite interactions are taken into account, the two
configurations are degenerate. In the case of $m$ extra bosons, the
degeneracy is $\frac{ N! }{ m! (N-m)! }$. The interbosonic distance is
different for each degenerate configuration. Since there is no
preferred distance, the extra bosons will not be initially trapped. If
the well depth keeps increasing, the extra bosons will eventually be
trapped too. But they will do so with many different wavevectors
${\vec q}_i$. The structure formed will be an amorphous solid.

In real life, besides on site interactions, there are interactions
acting at long distances. They can be attractive or repulsive. These
interactions are weaker than the on site interaction $U$, hence not
important for a commensurate system. But if there is incommensuration,
these interactions will probably play a role in determining the
positions of the extra bosons. Another possibility is that when
increasing $U$ the extra bosons get trapped in one of the possible
configurations. If the configurations are metastable, then the system
will remain there for a long time.

In our previous work, we had determined that a Mott localization
transition could happen for incommensurate values of the number of
bosons per site \cite{us}. The nature of what this phase was was not
elucidated. The difficulty lied in where to place the extra bosons
after the integer part was subtracted from the total. As far as the
Mott phase was concerned, they could fall in the wells at random as
long as not two of them lie in the same well, for this would cost
extra $U$'s of energy. In real life, an extra interaction might cause
them to order in a certain way and decide the positions where they
lie. We do not consider such a case here. Another possibility in the
Bose-Hubbard model is that the integer part of the number of bosons
per site (which is commensurate by definition) gets localized in each
well and the leftover fraction constitute a Bose fluid. This would
correspond to Mott-superfluid coexistence. We find that no such
coexistence is possible.

When the total number of bosons is very near a commensurate number,
the few leftover bosons are not localized and gain kinetic energy by
being spread through the lattice. On the other hand, we do not think
this constitutes a phase where we have coexistence of two phases. This
is because the fraction of bosons which can be spread over the lattice
collapses to zero as the lattice is made larger and larger. A true
thermodynamic phase survives the thermodynamic limit. In fact, its
properties become exact in the thermodynamic limit. Hence, the region
where the fraction of bosons is free over the lattice exists only due
to finite size effects. We thus conclude that for incommensurate
values there is a Mott phase where the fraction of bosons that causes
the incommensuration gets localized at random places in the lattice,
forming a kind of amorphous solid.

We now estimate how this comes about. We will study, making mean field
estimations, whether it is energetically favorable for the ground
state to have all the bosons localized, or to have a commensurate
number localized with the leftover fraction forming a gas or fluid
coexisting with the Mott phase. The coexisting fluid will of course be
a superfluid because of the repulsion, but we will only consider the
fluid to be a Bose condensed free boson gas as a precondition for
superfluidity is to have a Bose condensate.

In both the Mott phase with all the bosons localized and the phase
with coexistence of the BEC and the Mott insulator (MI), we have
$\sum_i \langle n_i \rangle = M$. Therefore, the chemical potential
term will not favor any of the phases. In this section it will be
dropped from our Bose Hubbard Hamiltonian for energetic comparisons:
\beq
\sum_{<ij>} -t \left( a_i^\dagger a_j + a_j^\dagger a_i \right) + U
\sum_i n_i^2 \; .
\eneq
\noindent Without loss of generality we will consider the case where
$M$ lies between $N$ and $2N$, which we parametrize as $M=(1+p)N$ with
$0 \le p \le 1$. The wavefunction that describes the Mott phase with
all particles localized is given by
\beq
|\Psi_M\rangle = \prod_{i=1}^M a_{j_i}^\dagger |0\rangle
\eneq
\noindent where the $j_i$'s run over lattice sites and the lattice
sites are filled such that there is not less than one and not more
than two particles per site. This way we have the minimum Hubbard
energy possible. The expected value of the kinetic energy in this
state is zero. Hence, the expected energy of this state is
\beq
E_{GS}^M = (3M-2N)U = UN(1+3p) \; .
\eneq
\noindent This follows straightforwardly because we have $M-N$ sites
with two particles at an energy cost $4U$, and $2N-M$ sites with one
particle at an energy cost $U$.

For the phase consisting of the coexistence of a commensurate Mott
phase ($N$ particles with one particle per well) with a BEC of the
leftover $M-N$ particles the wavefunction is
\beq
|\Psi_{MB}\rangle = (a_0^\dagger)^{M-N} \prod_{i=1}^N a_i^\dagger |0\rangle
\eneq
\noindent where $a_0^\dagger$ is the operator that creates a boson
with momentum $\vec{k}=0$. $a_0^\dagger = (1/\sqrt{N})\sum_i
a_i^\dagger$. Only the fluid particles contribute to the expected
value of the kinetic energy, which is given by 
\beq
\langle E_k \rangle = -6 t (M-N) \; . 
\eneq
\noindent In order to estimate the expected value of the
interaction energy, we need to make a few estimates. The Mott
component of this coexistence phase has exactly one particle per
well. The fluid component, of course, has a fluctuating number of
particles per well, whose average is $(M-N)/N$ leading to
\beq
\langle n_i \rangle = 1 + \frac{M-N}{N} = \frac{M}{N} \; .
\eneq
\noindent In order to estimate the expected value of $n_i^2$, we
reckon the following way: the Mott component of the coexistence phase
does not have particle number fluctuations per site, leading to
$(\Delta n_i^2)_M = 0$. The BEC component of the coexistence phase has
fluctuations per site leading to typical fluctuations of a random
liquid (classical or quantum) $(\Delta n_i^2)_{MB} = (M-N)^2/N$. This
number comes about because there is a quantum amplitude for {\it all}
the fluid particles to be at each well (this is what it means to be
fluid or delocalized). Hence, all the particles contribute to $\Delta
n_i^2$ such that $\Delta n_i^2 \propto (M-N)^2$. This needs to be
multiplied by the probability of finding a particle in the well, which
is $1/N$. The total fluctuations are given by quadrature: $\Delta n_i^2
= (\Delta n_i^2)_M + (\Delta n_i^2)_{MB}$. From this we immediately
obtain
\beq
\langle n_i^2 \rangle = \Delta n_i^2 + \langle n_i \rangle^2 =
\frac{(M-N)^2}{N} + \frac{M^2}{N^2} = p^2N + (1+p)^2 \; .
\eneq
\noindent The energy of the coexistence phase is 
\beq
E_{GS}^{MB} = -6 t Np + UN [(N+1)p^2 + 2p + 1] \; .
\eneq
\noindent We define the quantity 
\beq 
\frac{E_{GS}^{MB} - E_{GS}^M}{NU} = - \left( \frac{6t}{U} + 1 \right)p
+ (N+1)p^2
\eneq
\noindent which when positive, the Mott phase is stable and when
negative, the coexistence phase is stable. It follows that for $p$
between 0 and $((6t/U) + 1)/(N+1)$ the coexistence phase is stable
while for $p > ((6t/U) + 1)/(N+1)$ the incommensurate Mott phase is
stable.

We see that when $N$ bosons get localized into a Mott phase with a
single boson per site, the extra bosons one adds on top of these will
gain more kinetic energy by delocalizing than the potential energy
penalty coming from the overlap with the localized particles. Hence,
they form a sort of fluid. The potential energy cost becomes too much
when there is a fraction $p = ((6t/U) + 1)/(N+1)$. At this point they
will localize too. This will happen despite there being an
incommensurate number of them. In the absence of other long range
interactions to tell the extra bosons where to localize, they will
localize in random wells as long as no two of them go to the same
well. For macroscopic lattices we see that the fraction of bosons that
will delocalize becomes arbitrarily small. Therefore, this is a finite
lattice size effect and does not reflect the properties of a true
thermodynamic phase of matter, which by definition becomes better
defined in the macroscopic limit.

\section{Josephson vs. Individual Bosons Tunneling}

In order to clarify the difference between the Mott phase we have
uncovered and that of earlier work, we study BEC tunneling in an
optical lattice. It has been verified experimentally in one
dimensional optical lattices that there is coherent quantum tunneling
in BEC's\cite{becj}.  We will try to take an in-depth look at
tunneling in optical lattices in a general number of dimensions. Of
course, the more relevant cases to superfluid-Mott transitions are those of 
higher dimensions, as superfluid ordering cannot occur in one dimension (see 
Appendix \ref{1d}). In order for the tunneling to be relevant, the hopping or
tunneling energy scale, $t$, needs to be larger than the kinetic
energy scale for the whole lattice:
\beq
T_L = \frac{\hbar^2}{2 m \Omega^{2/3}}
\eneq
\noindent where $m$ is the mass of the bosons and $\Omega$ is the volume of 
the whole lattice. If the kinetic energy of the lattice overwhelms
the hopping, the tight-binding approximation with bosons hopping from lattice 
site to lattice site with tunneling $t$ is no longer valid for delocalized 
bosons. 

On dimensional grounds one expects the tunneling rate to be 
proportional to $t/\hbar$. On the other hand, the proportionality factor should
be a universal dimensionless function of the number of lattice sites $N$, the
number of bosons $M$ and/or the number of bosons in the condensate $M_0$, and 
the different energy scales in the problem. Besides $t$ and $\hbar^2 / 2 m 
\Omega^{2/3}$, there is the on-site repulsion at each well $U$, the well depth
$\varepsilon$, and the kinetic energy scale of each well:
\beq
T_w =\frac{\hbar^2}{2 m V^{2/3}}
\eneq
\noindent where $V$ is the volume of each well $\sim a^3$ with $a$ the
interlattice spacing. We, of course, have $N V = \Omega$. The wells
are taken as large boxes.  This approximation is valid as wells in
optical lattices are fairly large ($a \sim 100$'s of nm) in typical
lattices\cite{ari1}. 

The interwell tunneling rate or particle current is
\beq
J= \frac{t}{\hbar} F\left(M_0, N, \frac{t}{\varepsilon}, 
\frac{t}{U}, \frac{T_w}{U}\right)
\eneq
\noindent The universal function $F$ will behave differently in
different physical regimes. For example, if the wells are very deep,
$\varepsilon \gg t$, the tunneling rate should be exponentially
suppressed $t \sim \exp(- \alpha \varepsilon/t)$ with $\alpha$ a
positive numerical constant. We will never write the dependence of the
function $F$ on $t/\varepsilon$ because the only effect of
$\varepsilon$ for the suppression of tunneling is through the decrease of
$t$. The number of bosons in the condensate, $M_0$, is determined
uniquely by the number of bosons $M$, and the repulsion $U$. We see
that the relevant parameter that sets the scale for the interaction
for the BEC is the combination $UM_0/N$. Hence the universal
dimensionless function that determines the tunneling current
enhancement does not depend on $U, M_0$ and $N$ independently, but
rather
\beq 
F\left( M_0,N,\frac{t}{U},\frac{T_w}{U} \right) = F\left(
\frac{tN}{UM_0}, \frac{T_wN}{UM_0} \right)
\eneq

We remind the reader that the dispersion relation for the superfluid
is
\beq
E_{\vec k} \equiv \sqrt{\tilde{\epsilon}_{\vec
k}^2-\frac{4U^2M_0^2}{N^2}}=\frac{\hbar
k}{\sqrt{2m}}\sqrt{\frac{\hbar^2 k^2}{2m} + \frac{4UM_0}{N}}
\eneq
\noindent where we make the continuum approximation $\epsilon_{\vec
k}=\hbar^2 k^2/2m$. As $k \rightarrow 0$ the dispersion
becomes phononic, which leads to dissipationless flow. The dispersion
at large $k$ becomes $\hbar^2 k^2/2m$, which leads to dissipative
flow. Notice that the crossover between dissipative and nondissipative
flow happens at $k \simeq \sqrt{8mUM_0/\hbar^2N}$. This defines the
coherence length of the superfluid 
\beq\label{xiw}
\xi \simeq
\sqrt{\frac{\hbar^2N}{8mUM_0}} \;.
\eneq

When the system is probed at length scales $L > \xi$ it responds
collectively as a superfluid, while at length scales $L < \xi$ it
behaves like a gas of independent bosonic particles, thus responding
dissipatively, or even not behaving like a fluid at all if the spacing between 
levels is relevant. We note that an infinite system would become superfluid
for any $U$ no matter how small, as one can always go to a long enough length 
scale to see the system respond collectively as a superfluid. In here we do not
have in mind infinite systems but finite wells of superfluid. In this
case of finite systems, if the kinetic energy scale of the well, $T_w$, is 
greater than the effective interaction, $UM_0/N$, the system {\it will not} 
be a superfluid as the linear size of the system, $V^{1/3}$, is less than
the coherence length of the superfluid.  Independent of whether the bosons are
superfluid in each well,  the system could still be a superfluid within the 
whole lattice.  

For the full lattice, when the kinetic energy scale is set by
tunneling between sites ($\epsilon_{\vec k}\simeq tk^2a^2$), the
coherence length is 
\beq\label{xil}
\xi \simeq a\sqrt{\frac{Nt}{4M_0U}} \; . 
\eneq
\noindent The size of the
lattice is, of course, $L=Na$ and the condition for superfluidity
($\xi < L$) is $U/t > 1/4M_0N$. We see that either for a very large
system ($N \rightarrow \infty$) or very large number of bosons 
($M_0 \rightarrow \infty$), it becomes arbitrarily easy to
become superfluid in the sense that an arbitrarily small $U$ will
order the boson fluid into a superfluid at low enough temperature. Conversely,
as we make the system small, or we reduce the number of bosons, it becomes 
harder to order into a superfluid, which will show up experimentally as 
increased dissipation rates. 

Even though a high enough $U$ can order the system 
into a superfluid, too high a $U$ will make it into a Mott 
insulator\cite{fisher,us}. The transition to the Mott insulator can be 
continuous\cite{fisher} or discontinuous\cite{us}. The continuous transition 
occurs when there is a superfluid\cite{fisher} in each well and $U$ prevents 
Josephson tunneling between wells. The continuous transition only happens for 
a commensurate number of bosons per site. When the bosons in the wells are not 
superfluid {\it within} wells the transition is discontinuous and happens 
irrespective of whether the number of bosons per site is integer or 
not\cite{us}. The second transition always happens when the number of bosons 
per site or $U/t$ are sufficiently small as in these cases $\xi >a$. In fact 
from our estimate (\ref{xil}) there cannot be a superfluid within each well 
when
\beq
\frac{Nt}{4M_0U} > 1 \; .
\eneq
\noindent For this case the Mott transition will be like ours if this condition
still holds at the critical value of $U/t$, otherwise the Mott transition will 
be that of suppression of Josephson tunneling\cite{fisher}.

We now proceed to consider the case when there is a superfluid within
each well. This will elucidate the microscopic physics of how onsite
repulsion suppresses Josephson tunneling, leading to the transition
studied by Fisher, {\it et. al.} \cite{fisher}. We will see that
commensuration plays no role in the suppression of tunneling. This
leads us to suspect that a first order transition can occur for this
case too contrary to the early work \cite{fisher}. The reason this was
not found out in that work is that they perform renormalization group
studies which can only access continuous transitions. Whether a
discontinuous transition can occur in this case is a matter for
further study.

We evaluate the current operator\cite{joseph}
\begin{align}
\nonumber \hat{J} &\equiv \frac{d}{dT}(N-\bar{N}) =
-\frac{i}{\hbar}\left[ N-\bar{N},H_t \right] \\ &=
-\frac{2it}{\hbar}\sum_{\vec k}\left( a_{\vec k}^\dagger\bar{a}_{\vec
  k} - \bar{a}_{\vec k}^\dagger a_{\vec k} \right)
\end{align}
\noindent where 
\beq \label{tunham}
H_t=-t\sum_{\vec k}\left( a_{\vec k}^\dagger\bar{a}_{\vec k} +
\bar{a}_{\vec k}^\dagger a_{\vec k}\right)
\eneq
\noindent is the tunneling part of the Hamiltonian and $T$ denotes
time. Now $a_{\vec k}$ and $\bar{a}_{\vec k}$ are boson operators for different
adjacent wells. At first we consider the tunneling current when we have a
superfluid within each well. We note that the ground state of the
reduced Hamiltonian (\ref{redham}) when $t=0$ is an outer product of
the Bogolyubov ground states for each well, which we denote as our
``vacuum'' state $|\psi_0\rangle$. The excited states of such a system
are denoted by $|n\rangle$ and are constructed by applying any number
of Bogolyubov creation operators to the vacuum state. The energies of
such states, $E_n$, will be the sum of the energies of each of the
operators. When $t \neq 0$ there will be tunneling between adjacent
wells, which can be calculated accurately from linear response theory
\cite{pines} as long as the number of atoms within the wells is
large. For such a calculation the only relevant excited states are two
quasiparticle states, one in each adjacent well, with opposite momenta,
i.e. $|n\rangle=\bar{b}_{\vec k}^\dagger b_{-\vec k}^\dagger
|\psi_0\rangle$, with energies $E_n=2E_{\vec k}$. This calculation was
first done by B. D. Josephson for Cooper paired fermionic
superconductors\cite{joseph}. We will follow his calculation
closely. When the wells are superfluid, the tunneling current is
dominated by coherent tunneling of Bogolyubov correlated pairs.

In order to calculate in linear response theory\cite{pines}, we turn
on $H_t$ adiabatically from time $T=-\infty$, starting from the ground
state of the noninteracting system at such a time. If we expand the
wavefunction in terms of the noninteracting eigenstates
\beqs
|\Psi\rangle=\sum_n a_n(T)\exp{(-iE_nT/\hbar)}|n\rangle
\eneqs
\noindent we obtain the well known formula
\beq
a_n(T)=\delta_{n,0}-i\exp{(iE_nT/\hbar)}\langle
n|H_t|\psi_0\rangle/(iE_n + \epsilon) 
\eneq
\noindent where the vacuum state energy $E_0$ has been chosen to be 0.
The parameter $\epsilon$ was introduced to control the adiabatic
switching of the interaction. The causal limit $\epsilon \rightarrow
0^+$ should be always understood. Our interaction is turned on fully
at time $T=0$ and the wavefunction of the interacting system is given
by $|\Psi(0)\rangle = |\psi_0\rangle + |\delta\varphi\rangle$
where\cite{pines}
\beq
|\delta\varphi\rangle = -\sum_{n\neq 0}\frac{|n\rangle\langle
 n|H_t|\psi_0\rangle}{E_n - i\epsilon}
\eneq

The Josephson current is given by
$\langle\Psi(0)|\hat{J}|\Psi(0)\rangle=\langle0|\hat{J}|\delta\varphi\rangle+$
complex conjugate. With
\beq
|\delta\varphi\rangle=t\sum_{\vec k}\frac{u_{\vec k}v_{\vec
 k}}{2E_{\vec k}-i\epsilon} \left( e^{-i\theta}b_{\vec k}^\dagger
 \bar{b}_{-\vec k}^\dagger + b_{-\vec k}^\dagger \bar{b}_{\vec
 k}^\dagger\right) \; ,
\eneq
\noindent where $\theta$ is the phase difference between the
superfluid order parameters in adjacent wells, we obtain
\beq
J=-\frac{2t^2}{\hbar}\sin{\theta}\sum_{\vec k}\frac{u_{\vec
 k}^2v_{\vec k}^2}{E_{\vec k}}
\eneq
\noindent analogous to the fermionic superconducting
case \cite{joseph}. We note that if $\theta$ is 0 there is no Josephson
tunneling as the superfluids are in phase. Since $u_{\vec k}v_{\vec
k}=UM_0/E_{\vec k}N$, we obtain for the enhancement factor
$F=J\hbar/t$
\beq
F=-\frac{2tU^2M_0^2}{N^2}\sin{\theta}\sum_{\vec k}\frac{1}{E_{\vec k}^3}
\eneq
\noindent In order to estimate this quantity we approximate the
quasiparticle dispersion as phononic, $E_{\vec k}^2 \simeq
2UM_0\hbar^2 k^2/mN$, and we cut off the high momenta at $k_c$ of
the order of the inverse coherence length. Converting the sum into an
integral
\beq 
\sum_{\vec k}\frac{1}{E_{\vec k}^3} \simeq \frac{1}{(2\pi)^3}\left(
\frac{mNV^{2/3}}{2UM_0} \right)^{3/2} \int_{k < k_c}
\frac{d^3\vec{k}}{k^3}
\eneq
\noindent Note that this integral is infrared divergent but the
divergence gets cut-off by the inverse linear well size $V^{1/3}$. We
finally obtain for the enhancement
\beq\label{f}
F=\frac{1}{16\pi^2}\frac{tN}{UM_0}\left( \frac{UM_0}{T_wN}
\right)^{3/2}\; \ln{\frac{4UM_0}{T_wN}}
\eneq

In order for the system to be superfluid at all within a well we must
have the linear dimensions of the well larger than the coherence
length, $V^{1/3} > \xi$, and there needs to be enough bosons in each
well as to have superfluid broken symmetry. Otherwise, {\it there will
not be} Josephson tunneling per se and the calculation of tunneling
{\it will have} to be done for noninteracting bosons trapped in
wells. This condition is $UM_0/T_wN > 1/4$ with smaller values
corresponding to a physically impossible regime. When the conditions
are such that there can be Josephson tunneling between sites,
intrawell Coulomb repulsion will suppress Josephson tunneling, leading
to a Mott insulating phase as studied originally \cite{fisher}. On the
other hand, when Josephson tunneling cannot occur from well to well
because the coherence length is longer than the well size, the Mott
transition will be single boson localization as uncovered by us
\cite{us}. The conditions in Mott transitions in optical lattice
experiments \cite{ib,ari} are such that the coherence length is too
long and the number of bosons too few as to support Josephson
tunneling between adjacent wells as there would not be a superfluid
within each well.

We concentrate on the case when the linear size of the wells is
too small ($V^{1/3}< \xi$). In this case the interaction cannot order
the system into a superfluid within each well (although it can
certainly be a superfluid within the whole lattice), and its only
effect is a renormalization of the mass of the bosons. Hence the
system is described by the Hamiltonian (\ref{tunham}) with $m$ the
renormalized mass of the bosons. There is no Josephson tunneling per
se as the system is not superfluid. On the other hand, there can be
coherent quantum tunneling of the (nonsuperfluid) boson liquid between
wells.

The Hamiltonian (\ref{tunham}) is diagonalized by the canonical
operators
\beq
A_{\vec k} = \frac{a_{\vec k} + \bar{a}_{\vec k}}{\sqrt{2}} \qquad
\bar{A}_{\vec k} = \frac{a_{\vec k} - \bar{a}_{\vec k}}{\sqrt{2}}
\eneq
\noindent to yield
\beq
\mathcal{H}= \sum_{\vec k} \left \{(\epsilon_{\vec k} -t)A_{\vec k}^\dagger
A_{\vec k} +  (\epsilon_{\vec k} + t)\bar{A}_{\vec k}^\dagger \bar{A}_{\vec k} 
\right \} \label{diagh}
\eneq
\noindent The Heisenberg equations of motion lead to the following time 
dependent operators $A_{\vec k}(T) = A_{\vec k}\exp[-i(\epsilon_{\vec k} -t)T
/ \hbar]$ and $\bar{A}_{\vec k}(T) = \bar{A}_{\vec k}\exp[-i(\epsilon_{\vec k}
+t)T / \hbar]$. The common phase factor $-i\epsilon_{\vec k} T/ \hbar$ will be 
dropped as only phase differences affect the physics. From these time dependent
operators one immediately obtains the operators into which $a_{\vec k}$ 
$\bar{a}_{\vec k}$ and their conjugates develop under Hamiltonian evolution:
\begin{align}
\nonumber a_{\vec k}(T) = \cos\left( \frac{tT}{\hbar}\right)a_{\vec k} 
+ i \sin\left( \frac{tT}{\hbar}\right)\bar{a}_{\vec k} \\ \nonumber
\bar{a}_{\vec k}(T) = \cos\left( \frac{tT}{\hbar}\right)\bar{a}_{\vec k} 
+ i \sin\left( \frac{tT}{\hbar}\right)a_{\vec k} \\ \nonumber
a_{\vec k}^\dagger (T) = \cos\left( \frac{tT}{\hbar}\right)a_{\vec k}^\dagger 
- i \sin\left( \frac{tT}{\hbar}\right)\bar{a}_{\vec k}^\dagger \\
\bar{a}_{\vec k}^\dagger (T) = \cos\left( \frac{tT}{\hbar}\right)
\bar{a}_{\vec k}^\dagger -i \sin\left(\frac{tT}{\hbar}\right)a_{\vec k}^\dagger
\end{align}
\noindent The ``barred'' and ``unbarred'' operators are orthogonal raising and 
lowering operators as unitary Hamiltonian evolution is a canonical 
transformation thus preserving the commutation relations of the operators at 
time zero.

We can now calculate the coherent tunneling current between two wells.
We take the right well to have $(N + M)/2$ bosons and the left well with 
$(N - M)/2$ bosons. The initial wavefunction is then
\beq
| \Psi (0)\rangle = \Big [\frac{(a_0^\dagger)^{(N + M)/2} 
(\bar{a}_0^\dagger)^{(N - M)/2} }{(N/2+M/2)!(N/2-M/2)!} \Big ] |\psi_0\rangle
\eneq
\noindent where the subscript $0$ means the smallest momentum state into which 
the bosons condensed. The wavefunction at time $T$, $| \Psi (T)\rangle$, has 
the exact same dependence on the operators $a_{0}^\dagger (T)$, 
$\bar{a}_{0}^\dagger(T)$ that $| \Psi (0)\rangle$ has on 
$a_{0}^\dagger$, $\bar{a}_{0}^\dagger$. In order to obtain the 
coherent tunneling current $J$ we need to evaluate 
$\langle \Delta N(T) \rangle \equiv \langle \Psi (T)| \Delta N | 
\Psi (T)\rangle$ with $\Delta N \equiv a_{0}^\dagger a_{0} - 
\bar{a}_{0}^\dagger \bar{a}_{0}$ counting the excess number of bosons in the 
right well. Using
\begin{align} \nonumber
\Delta N =  \cos\left(\frac{2tT}{\hbar}\right)(a_{0}^\dagger(T) a_{0}(T) - 
\bar{a}_{0}^\dagger (T) \bar{a}_{0}(T)) \\ + 2 i
\sin\left(\frac{2tT}{\hbar}\right) (\bar{a}_{0}^\dagger(T) a_{0}(T) - 
a_{0}^\dagger(T) \bar{a}_{0}(T))
\end{align}
\noindent we obtain the tunneling current 
\beq
J= \frac{d}{dT}\langle \Delta N(T) \rangle = \frac{2t}{\hbar}M \sin\left(
\frac{2tT}{\hbar}\right)
\eneq
\noindent where the periodicity is an artifact of having a two site
system. The proportionality of the tunneling current on the initial
number difference of bosons between the sites, $M$, is not an
artifact. This coherent quantum tunneling can and will be suppressed
by strong enough repulsion within the wells, leading to boson
localization within the wells.

Before concluding we will make some order of magnitude estimate of
when the Mott transition happens. Let us first consider the case when
the coherence length is smaller than the well size and there is a
macroscopic number of bosons per well such that there can be a
superfluid within each well. In this case, the Mott transition will be
suppression of Josephson tunneling as studied theoretically before the
advent of BECs \cite{fisher}. In this case, the energy scale
associated with tunneling is the Josephson tunneling energy
\beq
E_J= t \frac{1}{16\pi^2}\frac{t}{T_w}\left( \frac{UM_0}{T_wN}
\right)^{1/2}\; \ln{\frac{4UM_0}{T_wN}}
\eneq
\noindent where $4UM_0 \ T_wN > 1$ for Josephson tunneling to be
possible. When the average Hubbard repulsion energy per site, $E_U = U
(M_0/N)^2$, is larger than the Josephson energy the Hubbard
interaction prevents tunneling, making the system into a Mott
insulator. Since we expect the logarithm not to deviate much from
factors of order 1, we can estimate this condition to be about
\beq 
\left(\frac{t}{T_w}\right)^2 \left(
\frac{UM_0}{T_wN} \right )^{-1/2} \sim \frac{M_0}{N}
\eneq
\noindent For the case of long coherence length or few bosons per
site, the transition will be individual boson localization with
suppression of individual boson tunneling rather than Josephson
tunneling. The energy scale for individual boson tunneling is set
roughly by $t$. We would expect the transition to the Mott phase to
happen when this energy is swamp by the Hubbard interaction energy
\beq 
\frac{U}{t} \sim \frac{N}{M_0} 
\eneq
\noindent which is very roughly a condensate depletion condition. When
the lattice tunneling is more relevant than the intra-well kinetic
energy, $t > T_W$ and the number of bosons per site is of order 1, the
boson localization phase transition described in this work will
happen. On the other hand, if $t < T_W$ or the number of bosons per
site is considerably greater than one, one expects a Mott transition
with suppression of Josephson tunneling and superfluidity within each
well. If the number of bosons per site is considerably less than one,
one can probably not speak of either transition.

\section{Conclusion}

In BEC's in an optical lattice the Mott transition is induced by
making the wells on the lattice sites deeper ($\varepsilon$ is the
lattice depth) thus suppressing the tunelling amplitude $t$ between
near neighbor sites, i.e. making the bosons ``heavy''. Thus the
relevant ratio of on-site repulsion $U$ to hopping $t$ increases. As
this ratio increases, the superfluid order parameter increases as
shown in the figure \ref{odplot}. The increase of the order parameter
means that the superfluid phase is becoming more stable over the Bose
condensed gas. On the other hand, we also plot the number of particles
in the condensate. This number is becoming small as $U/t$
increases. As the condensate gets depleted the system is becoming
``less fluid'' and will solidify into a Mott insulating phase with a
well defined number of bosons per site. When the coherence length is
longer than the interwell spacing, or the number of bosons per well
too small, the Bogolyubov order parameter will be zero within each
well and in the quantum solid phase. There will be a {\it
discontinuous} jump in the superfluid response at the Mott
transition. This transition can happen regardless of commensuration
and the conditions in most lattice experiments are conducive to this
transition \cite{ib,ari}.

The physics of the transition we have described in the present work
differs considerably from the one proposed in the pioneering
theoretical work on Bose systems \cite{fisher} and thus requires some
comments. The first and, perhaps, most important difference is that we
do not consider a phase only model as it is not appropriate to the
Mott insulator at low densities since a superfluid order parameter
cannot survive within each well. In that original work, the phase of
the superfluid was disordered by the increasing repulsion thus leading
to a transition. In such a transition the solid would have a superfluid
order parameter within {\it each} well, but the phase of the order
parameter will have become randomized exactly analogous to what
happens in Josephson junction arrays when the charging energy is
sufficiently large. Such an insulating phase does not correspond to
the Mott insulator we studied here and cannot exist at small number of
bosons per site as one needs a macroscopic number of particles to have
a superfluid order parameter. It can also not exist when the coherence
length of the superfluid is longer than the well size. Finally, we
stress that the phase only model studied in the early work
\cite{fisher,ja} is a correct model of an array of Josephson coupled
superfluid systems and should work for bosons in an optical lattice
when the number of bosons per site is large enough, and the coherence
length short enough as to make superfluid order within each well
possible. Josephson coupled systems can easily be studied in an
optical lattice \cite{mark1,becj}. It will be extremely interesting to
see what the experimental phase diagram turns out to be.

\appendix

\section{Derivation of the Bogolyubov Hamiltonian}\label{bogoham}

The interaction part of the Hamiltonian can be written as
\beqs
\mathcal H_{int} = \frac{U}{N} \sum_{ {\vec k}, {\vec k}_2, {\vec q} }
\; a_{ {\vec k} + {\vec q} }^\dagger \; ( a_{ {\vec k}_2 - {\vec q}
}^\dagger a_{\vec k} + \delta_{ {\vec k}, {\vec k}_2 - {\vec q} } ) \;
a_{ {\vec k}_2 }
\eneqs
\beq \label{fulham}
\mathcal H_{int} = \frac{ U }{ N }\sum_{ {\vec k}, {\vec k}_2, {\vec
q} } \; a_{ {\vec k} + {\vec q} }^\dagger a_{ {\vec k}_2 - {\vec q}
}^\dagger a_{\vec k} a_{ {\vec k}_2 } + \frac{ U }{ N }\sum_{ {\vec
k}, {\vec q} } a_{\vec k}^\dagger a_{\vec k}
\eneq
\beqs
= \mathcal H_I + \frac{ U }{ N }\sum_{ {\vec k},{\vec q} } a_{\vec
k}^\dagger a_{\vec k}
\eneqs
\noindent Now we  start analyzing terms from $\mathcal H_I$. Separating from 
the momentum sums all the terms with $a_0$ or $a_0^\dagger$ and
replacing the values of these operators by $\sqrt{M_0}$ since the
condensate is macroscopically occupied and these operators behave like
c-numbers, we get
\begin{widetext}
\begin{multline}
\mathcal H_I = \frac{ U M_0^2 }{ N } + \frac{ 4 U M_0 }{ N }\sum_{
{\vec k} \neq 0 } a_{\vec k}^\dagger a_{\vec k} +
\frac{2 U \sqrt{M_0}}{N} \sum_{ {\vec k}_2 \neq 0,\vec q } \sum_{ {\vec q} \neq
0 } \; \left( a_{\vec q}^\dagger a_{ {\vec k}_2 - {\vec q} }^\dagger
a_{ {\vec k}_2 } + a_{ {\vec k}_2 }^\dagger a_{ {\vec k}_2 - {\vec q}
} a_{\vec q} \right) \\
+ \frac{ U M_0 }{ N } \sum_{ {\vec q} \neq 0 } \; \left( a_{\vec
q}^\dagger a_{ - {\vec q} }^\dagger + a_{ - {\vec q} } a_{\vec q}
\right ) + \frac{ U }{ N } \sum_{ {\vec k}_2 \neq 0,\vec{q} } \sum_{ {\vec k}
\neq 0,-\vec{q} } \sum_{\vec q} \; a_{ {\vec k} + {\vec q} }^\dagger a_{ {\vec
k}_2 - {\vec q} }^\dagger a_{\vec k} a_{ {\vec k}_2 }
\end{multline}
\end{widetext}
\noindent We concentrate on the universal ground state and low energy
properties of the superfluid. We thus neglect the terms with factors
of $\sqrt{M_0}$ or smaller since the macroscopic occupation of the
condensate will make them unimportant. We arrive then at the reduced
Bogolyubov Hamiltonian
\begin{multline}
\mathcal H_{bog} = \frac{ U M_0^2}{ N } + \sum_{\vec k \neq 0} 
\tilde{\epsilon}_{\vec k}\; a_{\vec k}^\dagger a_{\vec k}\\
+ \frac{ U M_0 }{ N } \sum_{ {\vec k} \neq 0 } \; 
( a_{\vec k}^\dagger a_{ - {\vec k} }^\dagger + a_{ - {\vec k} } a_{\vec k} )
\end{multline}
\noindent where 
\beq
 \tilde {\epsilon}_{\vec k}  = \epsilon_{\vec k} + (4UM_0/N) 
- \mu \; .
\eneq 
\noindent The extra term $ \frac{ U}{ N }
\sum_{ {\vec k}, {\vec q} }  a_{\vec k}^\dagger a_{\vec k} $
in equation (\ref{fulham}) can be written as $ U 
 \sum_{\vec k} a_{\vec k}^\dagger a_{\vec k}$.  In this way, it is seen that 
such a term just
renormalizes the chemical potential. It has therefore been absorbed
into $\mu$.

\section{Bosonic Superfluid Ground State}
\label{groundstate}

The manipulations reviewed in the present appendix are original to Bogolyubov
\cite{bog} and we follow closely K. Huang\cite{huang}
The ground state wavefunction for the Bogolyubov Hamiltonian (\ref{redham})
has the form
\beq \label{gsform}
| \Psi_0 \rangle = \sum_{n_1, m_1} \sum_{n_2, m_2} \; ... \; (c_{n_1, m_1} 
c_{n_2, m_2} \; ... \; )| n_1 m_1; \; n_2 m_2; \; ... \; \rangle
\eneq
\noindent where $n_i$ is the number of particles with momentum ${\vec k}_i$ and 
$m_i$ is the number of particles with momentum $-{\vec k}_i$. It is obviously 
the state with no excitations
\beqs
b_{\vec k} | \Psi_0 \rangle = ( u_{\vec k} a_{\vec k} + v_{\vec k} 
a_{ -{\vec k} }^\dagger ) | \Psi_0 \rangle = 0 
\eneqs
\noindent Let $c_{n, m} = c_{n_1, m_1}c_{n_2, m_2} \; ... \;$, 
$u = u_{\vec k}$, $v = v_{\vec k}$ and $| n, m \rangle = | n_1 m_1; \; n_2 m_2; 
\; ... \; \rangle$. Then 
\beq
\sum_{n, m} ( u \; c_{n,m} \sqrt{n} \; | n-1, m \rangle + v \; c_{n, m} 
\sqrt{m + 1} \; | n, m + 1 \rangle ) = 0
\eneq
\noindent Changing dummy variables $n \rightarrow n + 1$ in the first term of 
the sum and $m \rightarrow m - 1$ in the second term
\beq \label{csgs}
\sum_{n, m} ( u \; c_{n + 1,m} \sqrt{n + 1} + v \; c_{n, m - 1} \sqrt{m} ) \;
| n, m \rangle = 0
\eneq
\noindent For $n \neq m$ and $m = 0$ we have
\beqs
u \; c_{n + 1, m} \sqrt{n + 1} \; + \; v \; c_{n, m - 1} \sqrt{m} = 0 
\eneqs
\beqs
c_{n + 1, 0} = 0 \quad \text {for all $n$'s}
\eneqs
\noindent for m = 1
\beqs
u \; c_{n + 1, 1} \sqrt{ n + 1 } \; + \; v \; c_{n, 0} = u \; c_{n + 1, 1} \
sqrt{ n + 1 } = 0 
\eneqs
\beqs
c_{n + 1, 1} = 0
\eneqs
\noindent and so on for all $c_{n,m}$.  Thus we only need to consider $n = m$ 
in $c_{n, m}$. In that case we have from (\ref{csgs})
\beqs
u \; c_{m, m} \; + \; v \; c_{m - 1, m - 1} = 0 
\eneqs
\beqs
c_{m, m} = - \frac{v}{u} c_{m - 1, m - 1} = ( - \frac{v}{u} )^2 
c_{m - 2, m - 2} =  ( - \frac{v}{u} )^3 c_{m - 3, m - 3}
\eneqs
\beq
c_{m, m} = ( - \frac{v}{u} )^m c_{0, 0}
\eneq
\noindent Substituting this into (\ref{gsform}) we find
\beq
| \Psi_0 \rangle = \prod_{\vec k} \sum_{n = 0}^\infty  ( - \frac{v_{\vec k}}
{u_{\vec k}} )^n c_{0, 0} | n, n \rangle
\eneq
\noindent where we can write the state composed of n particles with momentum 
${\vec k}$ and n particles with momentum $- {\vec k}$ as
\beq
| n, n \rangle = \frac{ (a_{\vec k}^\dagger )^n }{ \sqrt{n!} } 
\frac{ (a_{ - {\vec k} }^\dagger )^n }{ \sqrt{n!} } \; | 0 \rangle
\eneq
\noindent hence
\beq
| \Psi_0 \rangle = \prod_{\vec k} \sum_{n = 0}^\infty  ( - \frac{v_{\vec k}}
{u_{\vec k}} )^n c_{0, 0} \frac{ (a_{\vec k}^\dagger )^n 
(a_{ - {\vec k} }^\dagger )^n }{ n! } \; | 0 \rangle 
\eneq
\beq
| \Psi_0 \rangle = \prod_{\vec k} c_{0, 0} 
e^{ - \frac{ v_{\vec k} }{ u_{\vec k} } a_{\vec k}^\dagger 
a_{ - {\vec k} }^\dagger } | 0 \rangle
\eneq
\noindent which is a coherent state. We need to find the normalization 
constant $c_{0, 0}$. Notice that
\beqs
\langle \Psi_0 | \Psi_0 \rangle = |c_{0, 0}|^2 \langle 0| e^{ \frac{-v_{\vec k}}
{ u_{\vec k} } a_{\vec k} a_{ -{\vec k} } } e^{ \frac{ -v_{\vec k} }{ u_{\vec k} } 
a_{\vec k}^\dagger a_{ -{\vec k} }^\dagger } | 0 \rangle
\eneqs
\begin{multline*}
\langle \Psi_0 | \Psi_0 \rangle = |c_{0, 0}|^2 \sum_{n, m} (-1)^{n + m} 
( \frac{ v_{\vec k} }{ u_{\vec k} } )^{n + m} \\
\times \frac{ \langle 0 | a_{\vec k}^n 
a_{ -{\vec k} }^n ( a_{\vec k}^\dagger )^m ( a_{ -{\vec k} }^\dagger )^m 
| 0 \rangle }{ n! m! }
\end{multline*}
\noindent and furthermore
\beqs
\langle 0 | a_{\vec k}^n a_{-\vec k}^n (a_{\vec k}^\dagger )^m 
( a_{-\vec k}^\dagger )^m | 0 \rangle = \delta_{n, m} \langle 0 | 
a_{\vec k}^n a_{-\vec k} ^n ( a_{\vec k}^\dagger )^n 
( a_{-\vec k}^\dagger )^n | 0 \rangle
\eneqs
\beqs
a_{-\vec k }^\dagger | 0 \rangle = \sqrt{1} |1, 0\rangle \qquad
( a_{-\vec k}^\dagger )^2 | 0 \rangle = \sqrt{1} \sqrt{2} |2, 0\rangle
\eneqs
\beqs
( a_{-\vec k}^\dagger )^n | 0 \rangle = \sqrt{ n! } |n, 0\rangle
( a_{\vec k}^\dagger )^n | 0 \rangle = \sqrt{ n! } |0, n\rangle
\eneqs
\beqs
a_{-\vec k} | 0 \rangle = 0 \qquad
a_{-\vec k} | n \rangle = \sqrt{n} |n - 1\rangle
\eneqs
\beqs
( a_{-\vec k} )^2 | n \rangle = \sqrt {n} \sqrt{n - 1} |n - 2\rangle 
\qquad ( a_{-\vec k} )^n | 0 \rangle = \sqrt{ n! } | 0 \rangle
\eneqs
\noindent Using the expressions above we find
\begin{multline*}
\langle 0 | a_{\vec k}^n a_{-\vec k}^n ( a_{\vec k}^\dagger )^n 
( a_{-\vec k}^\dagger )^n | 0 \rangle =  n! \langle 0 | a_{\vec k}^n 
a_{-\vec k}^n | n, n \rangle \\
= ( n! )^2 \langle 0 | 0 \rangle = ( n! )^2
\end{multline*}
\noindent which finally yields
\beqs
\langle \Psi_0 | \Psi_0 \rangle = |c_{0, 0}|^2 \sum_{n, m} 
\frac{ (-1)^{n + m} ( \frac{ v_{\vec k} }{ u_{\vec k} } )^{n + m} }{ n! m! } 
\delta_{n, m} ( n! )^2 
\eneqs
\beqs
= |c_{0, 0}|^2 \sum_{n} 
( \frac{ v_{\vec k}}{ u_{\vec k}} )^ { 2 n } = 1
\eneqs
\beq
|c_{0, 0}|^2 = \frac{ 1 }{ \sum_n ( \frac{ v_{\vec k} }{ u_{\vec k} }
  )^{ 2n } }
\eneq
\noindent Remembering that $\sum_n ( \frac{ v_{\vec k}^2 }{ u_{\vec
k}^2 } )^n = \frac{ 1 }{ 1 - v_{\vec k}^2 / u_{\vec k}^2 } = u_{\vec
k}^2$ we arrive at the expression $|c_{0, 0}|^2 = \frac{ 1 }{ u_{\vec
k}^2 }$. The normalized ground state wavefunction is then
\beq 
\qquad \quad |\Psi_0 \rangle = \prod_{\vec k} \frac{ 1 }{ u_{\vec k} }
e^{ -(v_{\vec k} / u_{\vec k}) a_{\vec k}^\dagger a_{-\vec k}^\dagger}
\; |0 \rangle
\eneq

\section{Universal Superfluid physics}\label{uni}

In the present appendix we review and emphasize some of
the hydrodynamic and universal properties of the superfluid phase as
realized in Helium and BECs.

\subsection{Rigidity and Superfluidity}

We first review the Landau argument \cite{landau} illustrating how the
lack of rigidity causes dissipation. Consider a Bose fluid at zero
temperature (a Bose condensate) with excitation spectrum given by
$\epsilon_{\vec p}$. Suppose the fluid is moving with velocity
$\mathbf v$ and interacts with something external (what does it
interact with is irrelevant, but one could think that it is a surface
over which it is moving).

The interaction can exchange energy and momentum with the fluid. The fluid has 
initial energy $M v^2 /2$ with $M$ the total mass of the fluid. The interaction
creates a quasiparticle with energy $\epsilon_{\mathbf p}$ {\it in the frame} 
{\it of the Bose fluid.} In the lab frame the particle has energy
\beq
\tilde{\epsilon}_{\vec p} = \epsilon_{\vec p} + \vec p \cdot \vec v
\eneq
  \noindent by Galilean invariance since the velocity of the particle in the 
fluid frame is $\vec v_{\vec p} = \partial \epsilon_{\vec p} / 
\partial \vec p$ and it is $\vec v_{\vec p} + \vec v$ in the lab 
frame.

Energy will be exchanged if the fluid can lower its energy. The fluid final 
energy is
\beq \label{gali}
\frac{M v^2}{2} + \epsilon_{\vec p} + \vec p \cdot \vec v \; 
\eneq
\noindent The condition for dissipation is thus $\epsilon_{\vec p} + 
\vec p \cdot \vec v < 0$. The easiest way to satisfy the condition is 
when $\vec p$ is antiparallel to $\vec v$. We then obtain dissipation for
\beq
v > \left( \frac{\epsilon_{\vec p}}{p} \right)_{\text{min}} \; 
\eneq
\noindent We see that for all dispersions softer than sound the
minimum on the right hand side of the last equation is zero and there
is dissipation for {\it any velocity of the fluid.} If the dispersion
is at least as stiff as sound, for velocities smaller than some
critical velocity there is no dissipation and hence superflow. Since
as long as the system is a quantum Bose fluid, i.e. there exists a
condensate, repulsion produces sound excitations as the elementary low
energy quasiparticles of the system, the interactions provide the
necessary rigidity for superfluidity. The free Bose gas, despite the
macroscopic coherent occupation of the lowest momentum state, is not a
superfluid as it is arbitrarily easy to destroy its coherence no
matter {\it how weak the perturbation}. In fact, at energy scales in
which the superfluid behaves as a free Bose gas, it dissipates and it
is not super at all. This happens for wavelengths shorter than the
coherence length of the superfluid as explained in a subsequent
section.

\subsection{Scattering}

The purpose of this section is to illustrate the principles of
superfluidity and of macroscopic exactness by an example. Among the
experimental efforts on BECs figure studies of spontaneous emission
and light scattering of these superfluid systems by W. Ketterle and
collaborators\cite{kett1}. They found that light scattering off BECs
is suppressed to negligible values at long wavelengths. This seems
surprising at first since one can naively expect the typical $n + 1$
bosonic enhancement\cite{bec4} as observed in their emission
experiments\cite{kett1}. They have attributed the observed suppression
of light scattering in BECs to destructive quantum interference
effects\cite{bec4}. While their argument is certainly true, it is the
purpose of this section to point out that the more universal reason
that makes their argument, and hence their results, true is the {\it
Principle of Superfluidity}. We show that the suppression of light
scattering off a Bose Einstein Condensate is equivalent to the Landau
argument for superfluidity and thus is a consequence of the Principle
of Superfluidity.  This means that the superfluidity is ultimately the
reason for suppressed scattering at low wavelengths. We have
previously presented these arguments in a short note\cite{nos}.

We will study next scattering processes in the superfluid. In a scattering 
event, an incoming particle couples to the density of the system. The 
effective scattering Hamiltonian is
\beq
\mathcal H_I = \int d^3 r V ( {\bf r}, t ) \rho ({\bf r}) + \text{H.C.} 
\eneq
\noindent where by Fourier transforming
\beq
V ( {\bf r}, t ) = \sum_{\vec q} V_{\vec q} e^{-i ( {\vec q} \cdot {\bf r} 
- \omega t ) }
\eneq
\noindent Hence
\beq
\mathcal H_I = \sum_{\vec q} V_{\vec q} \int d^3 r \; 
e^{-i ( {\vec q} \cdot {\bf r} - \omega t ) } \rho ({\bf r}) + \text{H.C.} 
\eneq
\noindent $V_{\vec q}$ is the interaction matrix element between the
final and initial states of the external particle. Now, $\rho_{\vec q}
= \int d^3 r \; \rho ({\bf r}) e^{ -i {\vec q} \cdot {\bf r} }$ and at
an arbitrary point $i$, the density is given by $\rho ( {\bf r}_i ) =
a_i^\dagger a_i$, where
\beq
a_i = \sum_{\vec k} e^{ -i {\vec k} \cdot {\bf r} } a_{\vec k} \qquad 
a_i^\dagger = \sum_{ {\vec k}' } e^{ i {\vec k}' \cdot {\bf r} } 
a_{ {\vec k}' }^\dagger
\eneq
\noindent thus
\beq \label{rhor}
\rho ( {\bf r} ) = \sum_{ {\vec k}, {\vec k}' } e^{ -i ( {\vec k} -
{\vec k}' ) \cdot {\bf r} } a_{ {\vec k}' }^\dagger a_{\vec k}
\eneq
\begin{multline} \label{rhoq}
\int d^3 r \; \rho ({\bf r}) e^{ -i {\vec q} \cdot {\bf r} } = 
\sum_{ {\vec k}, {\vec k}' } a_{ {\vec k}' }^\dagger a_{\vec k} \int 
e^{ -i ( {\vec q} + {\vec k} - {\vec k}' ) \cdot {\bf r} } \; d^3 r \\
= \sum_{ {\vec k}, {\vec k}' } a_{ {\vec k}' }^\dagger a_{\vec k} 
\; \delta ( {\vec q} + {\vec k} - {\vec k}' ) = 
\sum_{\vec k} a_{ {\vec k} + {\vec q} }^\dagger a_{\vec k}
\end{multline}
\noindent and 
\beq
\mathcal H_I = \sum_{\vec k} V_{\vec q} e^{i \omega t}
a_{{\vec k} + {\vec q}}^\dagger a_{\vec k} + \text{H.C.}
\eneq
\noindent where the energy and momentum transfered in the scattering event are
\beq
\vec q = \vec q_f - \vec q_i \; \; \text {and} \; \; \omega = \omega_i - 
\omega_f
\eneq
\noindent and where we have concentrated only on one $\vec q$. Since at
first order the amplitudes for the different momenta add, it is enough
to do the calculations for one ${\vec q}$. This will not be true for
higher orders. Notice also that in the case of elastic scattering $\omega = 0$.

We need to insure adiabaticity in the interaction. Otherwise, a sudden hit on 
the system caused by an abrupt appearance of the scattering particle will do 
violent things to the system and cause it to generate entropy and heat up. We 
achieve this by the introduction of an extra term in the exponential of the 
interaction Hamiltonian\cite{pines}. The interaction will be turned on at 
$t = - \infty$ 
and left on until a time $t$. At the end of the calculations, we will take the 
limit $\eta \rightarrow 0$
\beq
\mathcal H_I = \sum_{\vec k} V_{\vec q} e^{i ( \omega - i \eta ) t }
a_{{\vec k} + {\vec q}}^\dagger a_{\vec k} + \text{H.C.}
\eneq

Calculating the time evolution of the Bose system in the presence of
the scattering particle by means of time dependent perturbation theory
we obtain for the amplitude
\beq
c_m (t) = \delta_{m, i} - \frac{ i }{ \hbar } \sum_n \delta_{n, i} 
\int_{-\infty}^t \langle \Psi_m | \mathcal H_I | \Psi_n \rangle 
e^{ i ( E_m - E_n ) t / \hbar } \; dt
\eneq
\noindent where in our case the initial state is the ground state: 
$| \Psi_i \rangle = | \Psi_o \rangle$, $E_i = 0$. For the scattering 
amplitude we are interested in $m \neq i$, where
\begin{multline}
\int_{-\infty}^t \langle \Psi_m | \mathcal H_I | \Psi_n \rangle 
e^{ i ( E_m - E_n ) t / \hbar } \; dt \\
= \sum_{\vec k} V_{\vec q} \langle \Psi_m | a_{ {\vec k} + {\vec q} }^\dagger 
a_{\vec k} | \Psi_n \rangle \int_{-\infty}^t e^{ i ( E_m - E_n 
+ \hbar \omega - i \hbar \eta ) t / \hbar } \; dt \\
= -i \hbar \sum_{\vec k} V_{\vec q} \frac{ \langle \Psi_m | 
a_{ {\vec k} + {\vec q} }^\dagger a_{\vec k} | \Psi_n \rangle }{ E_m - E_n 
+ \hbar \omega - i \hbar \eta } e^{ i ( E_m - E_n + \hbar \omega 
- i \hbar \eta ) t / \hbar }
\end{multline}
\noindent and thus
\begin{multline}
c_m (t) = - \sum_{\vec k} V_{\vec q} \frac{ \langle \Psi_m | 
a_{ {\vec k} + {\vec q} }^\dagger a_{\vec k} | \Psi_0 \rangle }{ E_m  
+ \hbar \omega - i \hbar \eta } e^{ i ( E_m + \hbar \omega - i \hbar \eta ) 
t / \hbar } \\
- \sum_{\vec k} V_{\vec q} \frac{ \langle \Psi_m | 
 a_{\vec k}^\dagger a_{ {\vec k} + {\vec q} } | \Psi_0 \rangle }{ E_m  
- \hbar \omega - i \hbar \eta } e^{ i ( E_m - \hbar \omega - i \hbar \eta ) 
t / \hbar }
\end{multline}
\noindent The scattering transition amplitude is, after separating 
the ${\vec k} = 0, -{\vec q}$ momentum states to account for the condensate
\begin{widetext} \label{scaamp}
\begin{multline}
c_m (t) = - \frac{ V_{\vec q} \sqrt {M_0} ( \langle \Psi_m | 
a_{\vec q}^\dagger | \Psi_0 \rangle + \langle \Psi_m | a_{-\vec q} 
| \Psi_0 \rangle ) }{E_m + \hbar \omega - i \hbar \eta } \; 
e^{i(E_m + \hbar \omega - i \hbar \eta) t/ \hbar} \; - \frac { V_{\vec q} 
\sqrt {M_0} ( \langle \Psi_m | a_{\vec q} |\Psi_0 \rangle + \langle \Psi_m 
| a_{-\vec q}^\dagger | \Psi_0 \rangle ) }{E_m - \hbar \omega - i \hbar \eta } 
\; e^{i(E_m - \hbar \omega - i \hbar \eta) t/ \hbar} \\
- \sum_{ {\vec k} \neq 0, -{\vec q} } V_{\vec q} \frac{ \langle \Psi_m | 
a_{ {\vec k} + {\vec q} }^\dagger a_{\vec k} | \Psi_0 \rangle }{ E_m  
+ \hbar \omega - i \hbar \eta } e^{ i ( E_m + \hbar \omega - i \hbar \eta ) 
t / \hbar } - \sum_{ {\vec k} \neq 0, -{\vec q} } V_{\vec q} 
\frac{ \langle \Psi_m | a_{\vec k}^\dagger a_{ {\vec k} + {\vec q} } 
| \Psi_0 \rangle }{ E_m  - \hbar \omega - i \hbar \eta } e^{ i ( E_m 
- \hbar \omega - i \hbar \eta ) t / \hbar }
\end{multline}
\end{widetext}

Let us look more closely at the first two terms. They represent
processes in which the external particle hits the atoms in the ground
state and tries to excite them, i.e. knock an atom into or knock it
out of the condensate. On the other hand, the entanglement in the
ground state makes sound the only possible excitation at low
momenta. In light of this, it is not too surprising that scattering is
suppressed. It is just the principle of superfluidity at work.

Since the scattering particles couple to the density, the amount of scattering 
is proportional to the density component with momentum $\vec q$, the so called 
density response function
\beq \label{density}
\langle \rho_{\vec q} \rangle = \sum_m c_m \langle \Psi_0 | \rho_{\vec q} |
\Psi_m \rangle + \sum_m c_m^* \langle \Psi_m | \rho_{\vec q} | \Psi_0 
\rangle
\eneq
\noindent Notice from (\ref{bogo}) that 
\beq
a_{\vec q}^\dagger = u_{\vec q} b_{\vec q}^\dagger - v_{\vec q} b_{ -
{\vec q} } \qquad a_{\vec q} = u_{\vec q} b_{\vec q} - v_{\vec q} b_{
- {\vec q} }^\dagger
\eneq
\noindent Using these relations together with (\ref{rhor}) and
(\ref{rhoq}) we see that
\beq \label{0densm}
\langle \Psi_0 | \rho_{\vec q} | \Psi_m \rangle = \sum_{\vec k}
\langle \Psi_0 | a_{ {\vec k} + {\vec q} }^\dagger a_{\vec k} | \Psi_m
\rangle e^{ - E_m t / \hbar }
\eneq
\noindent where 
\begin{widetext}
\begin{align}
\nonumber \langle \Psi_0 | a_{ {\vec k} + {\vec q} }^\dagger a_{\vec k} |
\Psi_m \rangle &= \langle \Psi_0 | ( u_{ {\vec k} + {\vec q} } b_{ {\bf
k} + {\vec q} }^\dagger - v_{ {\vec k} + {\vec q} } b_{ - {\vec k} - {\bf
q} } ) ( u_{\vec k} b_{\vec k} - v_{\vec k} b_{ - {\vec k} }^\dagger ) |
\Psi_m \rangle = - v_{ {\vec k} + {\vec q} } u_{\vec k} \langle \Psi_0 |
b_{ - {\vec k} - {\vec q} } b_{\vec k} | \Psi_m \rangle \\ 
\nonumber &+ \; v_{ {\vec k} + {\vec q} } v_{\vec k} \langle \Psi_0 |
b_{ - {\vec k} - {\vec q} } b_{ - {\vec k} }^\dagger | \Psi_m \rangle
= - v_{ {\vec k} + {\vec q} } u_{\vec k} \langle \Psi_0 | b_{ - {\vec
k} - {\vec q} } b_{\vec k} | \Psi_m \rangle \; + \; v_{ {\vec k} +
{\vec q} } v_{\vec k} \langle \Psi_0 | ( \delta ( {\vec q} ) + b_{ -
{\vec k} }^\dagger b_{ - {\vec k} - {\vec q} } |\Psi_m \rangle \\
&= - v_{ {\vec k} + {\vec q} } u_{\vec k} \langle \Psi_0 | b_{ - {\vec
k} - {\vec q} } b_{\vec k} | \Psi_m \rangle \; + \; v_{ {\vec k} +
{\vec q} } v_{\vec k} \langle \Psi_0 | \delta ( {\vec q} ) | \Psi_m
\rangle = - v_{ {\vec k} + {\vec q} } u_{\vec k} \langle \Psi_0 | b_{
- {\vec k} - {\vec q} } b_{\vec k} | \Psi_m \rangle
\end{align}
\end{widetext}
\noindent Since we are considering scattering with nonzero $\vec q$,
$\delta ( \vec q ) = 0$. The case where $ \vec q = 0 $
corresponds to no momentum transfer, i.e. no scattering and we discard it. 
Then the only final state giving a nonzero matrix element is
\beq \label{psim1}
| \Psi_{m_1} \rangle = b_{ -\vec k - \vec q }^\dagger b_{\vec k}^\dagger | 
\Psi_0 \rangle
\eneq
\noindent thus giving
\beq \label{matriz0m2}
\langle \Psi_0 | a_{ {\vec k} + {\vec q} }^\dagger a_{\vec k} | \Psi_m
\rangle = - v_{ {\vec k} + {\vec q} } u_{\vec k}
\eneq
\noindent For ${\vec k} = 0$ we have similarly
\begin{align}
\nonumber \langle \Psi_0 | \rho_{\vec q} | \Psi_m \rangle &= \sqrt{ M_0 } 
\langle \Psi_0 | a_{\vec q}^\dagger | \Psi_m \rangle \\
\nonumber &= \sqrt{ M_0 } \langle \Psi_0 | ( u_{\vec q} b_{\vec q}^\dagger 
- v_{\vec q} b_{ - {\vec q} } ) | \Psi_m \rangle \\
&= - \sqrt{ M_0 } v_{\vec q} \langle \Psi_0 | b_{ - {\vec q} } | 
\Psi_m \rangle 
\end{align}
\noindent In this case the only final state not killing the matrix element is 
\beq \label{psim0}
| \Psi_{m_0} \rangle = b_{ - {\vec q} }^\dagger | \Psi_0 
\rangle
\eneq
\noindent which gives
\beq \label{matriz0m1}
\langle \Psi_0 | a_{\vec q}^\dagger | \Psi_m \rangle  = - v_{\vec q}
\eneq
\noindent The same calculation for ${\vec k} = -{\vec q}$ yields
\begin{align}
\nonumber \langle \Psi_0 | \rho_{\vec q} | \Psi_m \rangle &= \sqrt{ M_0 } 
\langle \Psi_0 | a_{ - {\vec q} } | \Psi_m \rangle \\
\nonumber &= \sqrt{ M_0 } \langle \Psi_0 | ( u_{\vec q} b_{ - {\vec q} } 
- v_{\vec q} b_{\vec q}^\dagger ) | \Psi_m \rangle \\
&= \sqrt{ M_0 } u_{\vec q} \langle \Psi_0 | b_{ - {\vec q} } | 
\Psi_m \rangle 
\end{align}
\noindent with the same final state (\ref{psim0}). Continuing the
same type of calculation we also find
\begin{widetext}
\begin{align}
\nonumber \langle \Psi_m | a_{ \vec k + \vec q }^\dagger a_{\vec k} | 
\Psi_0 \rangle &= \langle \Psi_m | ( u_{ \vec k + \vec q } 
b_{ \vec k + \vec q }^\dagger - v_{ \vec k + \vec q } 
b_{ - \vec k - \vec q } ) ( u_{\vec k} b_{\vec k} - v_{\vec k} 
b_{ -\vec k }^\dagger ) | \Psi_0 \rangle = - u_{ \vec k + \vec q } 
v_{\vec k} \langle \Psi_m | b_{ \vec k + \vec q }^\dagger 
b_{ -\vec k }^\dagger | \Psi_0 \rangle \\
\nonumber &+ \; v_{ \vec k + \vec q } v_{\vec k} \langle \Psi_m | 
b_{ - \vec k - \vec q } b_{ - \vec k }^\dagger | \Psi_0 \rangle = 
- u_{ \vec k + \vec q } v_{\vec k} \langle \Psi_m | 
b_{ \vec k + \vec q }^\dagger b_{ - \vec k }^\dagger | \Psi_0 \rangle \; 
+ \; v_{ \vec k + \vec q } v_{\vec k} \langle \Psi_m | ( \delta ( \vec q ) + 
b_{ - \vec k }^\dagger b_{ - \vec k - \vec q } |\Psi_0 \rangle \\
&= - u_{ \vec k + \vec q } v_{\vec k} \langle \Psi_m | b_{ \vec k + 
\vec q }^\dagger b_{ - \vec k }^\dagger | \Psi_0 \rangle + \; 
v_{ \vec k + \vec q } v_{\vec k} \langle \Psi_m | \delta ( \vec q ) |
\Psi_0 \rangle = - u_{ \vec k + \vec q } v_{\vec k} \langle \Psi_m | 
b_{ \vec k + \vec q }^\dagger b_{ - \vec k }^\dagger | \Psi_0 \rangle
\end{align}
\end{widetext}
\noindent Again, since ${\vec q} \neq 0$ in our considerations, 
$\delta ( {\vec q} ) = 0$ and we find as the only possible final state
\beq \label{psim1'}
| \Psi_{m_1'} \rangle = b_{ {\vec k} + {\vec q} }^\dagger 
b_{ - {\vec k} }^\dagger | \Psi_0 \rangle
\eneq
\noindent corresponding to a matrix element
\beq \label{matrizm02}
\langle \Psi_m | a_{ \vec k + \vec q }^\dagger a_{\vec k} | \Psi_0 \rangle
= - u_{ \vec k + \vec q} v_{\vec k}
\eneq
\noindent For $\vec k = 0$
\begin{align} \label{m0k0}
\nonumber \langle \Psi_m | a_{\vec q}^\dagger | \Psi_0 \rangle &= 
\langle \Psi_m | ( u_{\vec q} b_{\vec q}^\dagger - v_{\vec q} 
b_{ - {\vec q} } ) | \Psi_0 \rangle \\
&= u_{\vec q} \; \langle \Psi_m | b_{\vec q}^\dagger | 
\Psi_0 \rangle 
\end{align}
\noindent with final state
\beq \label{psim0'}
| \Psi_{m_0'} \rangle = b_{\vec q}^\dagger | \Psi_0 \rangle
\eneq
\noindent and matrix element
\beq \label{matrizm01}
\langle \Psi_m | a_{\vec q}^\dagger | \Psi_0 \rangle = u_{\vec q}
\eneq
\noindent Finally, setting ${\vec k}$ = $-{\vec q}$ we obtain
\begin{align} \label{m0k-q}
\nonumber \langle \Psi_m | a_{ - {\vec q} } | \Psi_0 \rangle &= \langle 
\Psi_m | ( u_{ - {\vec q} } b_{ - {\vec q} } - v_{ - {\vec q} } 
b_{\vec q}^\dagger ) | \Psi_0 \rangle \\
&= - v_{\vec q} \; \langle \Psi_m | b_{\vec q}^\dagger | 
\Psi_0 \rangle 
\end{align}
\noindent whose final state is the same state (\ref{psim0'})

Substituting (\ref{psim1}) and (\ref{psim0}) into (\ref{scaamp}) we
find the two scattering amplitudes $c_{m_0}$ (which corresponds to
both $\vec k = 0$ and $\vec k = -\vec q$ since both have the same final
state) and $c_{m_1}$ (corresponding to processes with ${\vec k} \neq 0,
-{\vec q}$) to be
\beq \label{cm0}
c_{m_0} = \frac{ - V_{\vec q} \sqrt{ M_0 } ( u_{\vec q} - v_{\vec q} ) }
{ E_{\vec q} - \hbar \omega - i \hbar \eta } e^{ i ( E_{\vec q} - \hbar \omega 
- i \hbar \eta ) t / \hbar }
\eneq
\beq \label{cm1}
c_{m_1} = \frac{ V_{\vec q} u_{\vec k} v_{ {\vec k} + {\vec q} } }{
E_{ {\vec k} + {\vec q} } + E_{\vec k} - \hbar \omega - i \hbar \eta }
e^{ i ( E_{ {\vec k} + {\vec q} } + E_{\vec k} - \hbar \omega - i
\hbar \eta ) t / \hbar }
\eneq
\noindent On the other hand, substituting (\ref{psim1'}) and
(\ref{psim0'}) into (\ref{scaamp}) gives for the scattering
amplitudes $c_{m_0'}$ and $c_{m_1'}$
\beq
c_{m_0'} = \frac{ - V_{\vec q} \sqrt{ M_0 } ( u_{\vec q} - v_{\vec q} ) }
{ E_{\vec q} + \hbar \omega - i \hbar \eta } e^{ i ( E_{\vec q} + \hbar \omega 
- i \hbar \eta ) t / \hbar }
\eneq
\beq
c_{m_1'} = \frac{ V_{\vec q} v_{\vec k} u_{ {\vec k} + {\vec q} } }{
E_{ {\vec k} + {\vec q} } + E_{\vec k} + \hbar \omega - i \hbar \eta }
e^{ i ( E_{ {\vec k} + {\vec q} } + E_{\vec k} + \hbar \omega - i
\hbar \eta ) t / \hbar }
\eneq
\noindent hence
\beq \label{cm0'*}
c_{m_0'}^* = \frac{ - V_{\vec q} \sqrt{ M_0 } ( u_{\vec q} - v_{\vec
q} ) } { E_{\vec q} + \hbar \omega + i \hbar \eta } e^{ - i ( E_{\vec
q} + \hbar \omega + i \hbar \eta ) t / \hbar }
\eneq
\beq \label{cm1'*}
c_{m_1'}^* = \frac{ V_{\vec q} v_{\vec k} u_{ {\vec k} + {\vec q} } }{
E_{ {\vec k} + {\vec q} } + E_{\vec k} - \hbar \omega + i \hbar \eta }
e^{ - i ( E_{ {\vec k} + {\vec q} } + E_{\vec k} - \hbar \omega + i
\hbar \eta ) t / \hbar }
\eneq
\noindent Plugging all these results into (\ref{density}) we obtain
\begin{widetext}
\begin{align}
\langle \rho_{\vec q} \rangle &= \frac{ - V_{\vec q} M_0 
( u_{\vec q} - v_{\vec q} )^2 }{ E_{\vec q} - \hbar \omega - i \hbar \eta } 
e^{ i ( E_{\vec q} - \hbar \omega - i \hbar \eta ) t / \hbar } 
e^{ -i E_{\vec q} t / \hbar } - \frac{ V_{\vec q} M_0 
( u_{\vec q} - v_{\vec q} )^2 }{ E_{\vec q} + \hbar \omega + i \hbar \eta } 
e^{ -i ( E_{\vec q} + \hbar \omega + i \hbar \eta ) t / \hbar } 
e^{ i E_{\vec q} t / \hbar }  \nonumber \\
&- \sum_{m} \sum_{ \vec k \neq 0, -\vec q } 
\sum_{ \vec l \neq 0, -\vec q } \frac{ u_{\vec k} u_{\vec l} 
v_{ \vec k + \vec q } v_{ \vec l + \vec q } V_{\vec q}} 
{ E_{ \vec k + \vec q } + E_{\vec k} - \hbar \omega - i \hbar \eta } 
e^{ i ( E_{ \vec k + \vec q } + E_{\vec k} - \hbar \omega - i \hbar \eta ) 
t / \hbar } e^{ -i ( E_{ \vec k + \vec q } + E_{\vec k} ) t / \hbar } \\
&- \sum_{m} \sum_{ \vec k \neq 0, -\vec q } 
\sum_{ \vec l \neq 0, -\vec q} \frac{ v_{\vec k} v_{\vec l} 
u_{ \vec k + \vec q } u_{ \vec l + \vec q } V_{\vec q }}
{ E_{\vec k + \vec q } + E_{\vec k} + \hbar \omega + i \hbar \eta } 
e^{ -i ( E_{ \vec k + \vec q } + E_{\vec k} + \hbar \omega + i \hbar \eta ) 
t / \hbar } e^{ i ( E_{ \vec k + \vec q } + E_{\vec k} ) t / \hbar }
\nonumber
\end{align}
\end{widetext}

The higher momenta terms ($\vec k \neq 0, -\vec q$) in the density
response function represent processes that are forbidden by kinematics
constraints, hence they are dropped. In the case of light scattering
conservation of energy gives
\beq
c q_i = c q_f + 2 c_s k
\eneq
\noindent while conservation of momentum gives
\beq \label{econs}
\vec q_i = \vec q_f + \vec k  - \vec k
\eneq
\noindent for there is a particle excited with momentum $\vec k$ and
another one with momentum $-\vec k$. We have
\begin{align}
\nonumber \vec q_i &= \vec q_f \\
c \vec q_i &= c \vec q_f
\end{align}
\noindent which can only be satisfied simultaneously with the
conservation of energy condition in (\ref{econs}) by $\vec k =
0$. Thus the processes are forbidden.

As for the multiplicative factor $( u_{\vec q} - v_{\vec q} )^2$, it is
an interference factor which provides suppression of light scattering
as required by the necessary rigidity of the ground state. Expanding
it out and using equations (\ref{etilde}), (\ref{uv}),
(\ref{energy}), and (\ref{sound}) we find, as ${\vec q} \rightarrow
0$
\beq
( u_{\vec q} - v_{\vec q} )^2 = u_{\vec q}^2 + v_{\vec q}^2 - 2 u_{\vec q}
v_{\vec q}
\eneq
\beq
\tilde {\epsilon}_{\vec q} = M_0 U_0 ( 1 + \frac{ \hbar^2 q^2 }
{ 2 m M_0 U_0 } ) = M_0 U_0 ( 1 + \frac{ \hbar^2 q^2 }{ 2 m^2 c_s^2 } )
\eneq
\beq
E_{\vec q} = \hbar q \sqrt{ \frac{ M_0 U_0 }{ m } } \left( 1 + \frac{
\hbar^2 q^2 } { 4 m M_0 U_0 } \right)^{ 1/2 } \simeq \hbar q c_s
\left( 1 + \frac{ \hbar^2 q^2 } { 8 m^2 c_s^2 } \right)
\eneq
\noindent thus 
\begin{align}
\nonumber \frac{ \tilde {\epsilon}_{\vec q} }{ E_{\vec q} } &\simeq
\frac{ M_0 U_0 \left( 1 + \frac{ \hbar^2 q^2 }{ 2 m^2 c_s^2 } \right)
}{ \hbar q c_s \left( 1 + \frac{ \hbar^2 q^2 } { 8 m^2 c_s^2 } \right)
} \\ \nonumber 
&\simeq  \frac{ m c_s }{ \hbar q } \left( 1 + \frac{ \hbar^2 q^2 } { 2
m^2 c_s^2 } \right) \left( 1 - \frac{ \hbar^2 q^2 }{ 8 m^2 c_s^2 }
\right) \\
&\simeq \frac{ m c_s }{ \hbar q } \left( 1 + \frac{ 3 \hbar^2 q^2 }{ 8
m^2 c_s^2 } \right)
\end{align}
\beq
u_{\vec q}^2 = \frac{ 1 }{ 2 } \left( \frac{ \tilde {\epsilon}_{\vec
q} } { E_{\vec q} } + 1 \right) \qquad v_{\vec q}^2 = \frac{ 1 }{ 2
}\left( \frac{ \tilde {\epsilon}_{\vec q} } { E_{\vec q} } - 1 \right)
\eneq
\noindent Hence
\beq
u_{\vec q}^2 + v_{\vec q}^2 = \frac{ \tilde {\epsilon}_{\vec q} }
{ E_{\vec q} } \simeq \frac{ m c_s }{ \hbar q } \left( 1 + 
\frac{ 3 \hbar^2 q^2 }{ 8 m^2 c_s^2 } \right)
\eneq
\begin{align}
\nonumber u_{\vec q} v_{\vec q} &= \frac{ 1 }{ 2 } \sqrt{ \frac{
\tilde {\epsilon}_{\vec q}^2 }{ E_{\vec q}^2 } - 1 } \simeq \frac{ m
c_s }{ 2 \hbar q } \sqrt{\left( 1 + \frac{ 3 \hbar^2 q^2 }{ 8 m^2
c_s^2 } \right)^2 - \frac{ \hbar^2 q^2 }{ m^2 c_s^2 } } \\
&\simeq \frac{ m c_s }{ 2 \hbar q } \sqrt{ 1 - \frac{ \hbar^2 q^2 }{ 4
    m^2 c_s^2 } } \simeq \frac{ m c_s }{ 2 \hbar q } \left( 1 - \frac{
    \hbar^2 q^2 }{ 8 m^2 c_s^2 } \right)
\end{align}
\noindent As can be clearly seen from (\ref{m0k0}) and (\ref{m0k-q}),
$u_{\vec q}^2$ represents the probability of knocking a boson with
momentum $\vec q$ out of the condensate, $v_{\vec q}^2$ represents the
probability of knocking into the condensate one of the boson that are
correlated with momentum $-\vec q$, and $u_{\vec q} v_{\vec q}$ is the
interference between the two processes. Putting everything together
\begin{align}
\nonumber u_{\vec q}^2 + v_{\vec q}^2 - 2 u_{\vec q} v_{\vec q}
&\simeq \frac{m c_s}{\hbar q} \left( 1 + \frac{ 3 \hbar^2 q^2 }{ 8 m^2
c_s^2 } \right) - \frac{ m c_s }{ \hbar q } \left( 1 - \frac{ \hbar^2
q^2 }{ 8 m^2 c_s^2 } \right) \\
&= \frac{ \hbar q }{ 2 m c_s }
\end{align}
\noindent This goes to zero linearly as $\vec q \rightarrow
0$.  The processes of knocking a particle with momentum ${\vec q}$ out of the
correlated ground state and knocking a particle with momentum $-{\vec q}$
into the ground state both lead to the same final state.This is so because
neither of these is an elementary excitation of the system, but a
quantum superposition of them makes a Bogolyubov quasiparticle, which
is an eigenstate of the system. Different processes leading to the
same final state cannot be differentiated quantum mechanically and
thus will interfere. The interference is destructive at small
wavevectors and scattering is suppressed. More fundamentally, this
result follows from the rigidity concomitant to the superfluid state,
which is ultimately due to repulsive interactions between the
bosons. There is no scattering because the ``Principle of
Superfluidity'' forbids it. These are universal properties valid for
any superfluid state which will hold exactly at {\it low enough energy scales}.

We conclude by emphasizing that if there is no repulsion between the
atoms, low energy excitations will not have a sound spectrum, $v_{\vec
k} = 0, u_{\vec k} = 1$, leading to no superfluidity and no suppressed
light scattering. We thus see that suppressed light scattering follows
from the rigidity concomitant to the superfluid state, characterized
by $v_{\vec k} \neq 0$, which is equivalent to a nonzero sound
speed. This rigidity is ultimately due to repulsive interactions
between the bosons. There is no scattering because the ``Principle of
Superfluidity'' forbids it. These are universal properties valid for
any superfluid state which will hold exactly at {\it low enough energy
scales}.

\section{Some useful results}

\noindent We present the calculation of some operators in the superfluid 
ground state
\begin{align} \label{aasfgs}
\nonumber a_{ {\vec k}_1 } a_{ {\vec k}_2 } | \Psi_0 \rangle &= 
\frac{ \partial }{ \partial a_{ {\vec k}_1 }^\dagger } 
\frac{ \partial }{ \partial a_{ {\vec k}_2 }^\dagger } | \Psi_0 \rangle \\
\nonumber &= \frac{ \partial }{ \partial a_{ {\vec k}_1 }^\dagger } 
( \frac{ - v_{ {\vec k}_2} }{ u_{ {\vec k}_2 } } a_{ - {\vec k}_2 }^\dagger 
| \Psi_0 \rangle ) \\
\nonumber &= \frac{ - v_{ {\vec k}_2} }{ u_{ {\vec k}_2 } } 
\delta_{ {\vec k}_1 , - {\vec k}_2 } | \Psi_0 \rangle 
+ \frac{ v_{ {\vec k}_1} }{ u_{ {\vec k}_1 } }
\frac{ v_{ {\vec k}_2} }{ u_{ {\vec k}_2 } } 
a_{ - {\vec k}_1 }^\dagger a_{ - {\vec k}_2 }^\dagger | \Psi_0 \rangle \\
\nonumber &= \frac{ - v_{ {\vec k}_2} }{ u_{ {\vec k}_2 } } 
\delta_{ {\vec k}_1 , - {\vec k}_2 } | \Psi_0 \rangle \\
\nonumber & \qquad \qquad + \frac{ v_{ {\vec k}_1} }{ u_{ {\vec k}_1 } }
\frac{ v_{ {\vec k}_2} }{ u_{ {\vec k}_2 } }
( u_{ {\vec k}_1 } b_{ - {\vec k}_1 }^\dagger 
- v_{ {\vec k}_1 } b_{ {\vec k}_1 } ) \\
\nonumber & \qquad \qquad \times ( u_{ {\vec k}_2 } 
b_{ - {\vec k}_2 }^\dagger - v_{ {\vec k}_2 } b_{ {\vec k}_2 } ) 
| \Psi_0 \rangle \\
\nonumber &= \frac{ - v_{ {\vec k}_2} }{ u_{ {\vec k}_2 } } 
\delta_{ {\vec k}_1 , - {\vec k}_2 } | \Psi_0 \rangle 
+ \frac{ v_{ {\vec k}_1} }{ u_{ {\vec k}_1 } }
\frac{ v_{ {\vec k}_2} }{ u_{ {\vec k}_2 } } \\
\nonumber & \qquad \qquad \times ( u_{ {\vec k}_1 } b_{ - {\vec k}_1 }^\dagger 
- v_{ {\vec k}_1 } b_{ {\vec k}_1 } ) u_{ {\vec k}_2 } 
b_{ - {\vec k}_2 }^\dagger | \Psi_0 \rangle \\
\nonumber &= \frac{ - v_{ {\vec k}_2} }{ u_{ {\vec k}_2 } }
\delta_{ {\vec k}_1 , - {\vec k}_2 } | \Psi_0 \rangle \\
\nonumber & \qquad \qquad + v_{ {\vec k}_1} v_{ {\vec k}_2 } 
b_{ - {\vec k}_1 }^\dagger b_{ - {\vec k}_2 }^\dagger | \Psi_0 \rangle \\
\nonumber & \qquad \qquad - \frac{ v_{ {\vec k}_1 }^2 v_{ {\vec k}_2 } }
{ u_{ {\vec k}_1 } } b_{ {\vec k}_1 } 
b_{ - {\vec k}_2 }^\dagger | \Psi_0 \rangle \\
\nonumber &= \frac{ - v_{ {\vec k}_2} }{ u_{ {\vec k}_2 } }
( 1 + \frac{v_{{\vec k}_1}^2 u_{{\vec k}_2}}{u_{{\vec k}_1}} ) 
\delta_{ {\vec k}_1 , - {\vec k}_2 } | \Psi_0 \rangle \\
& \qquad \qquad + v_{ {\vec k}_1} v_{ {\vec k}_2 } 
b_{ - {\vec k}_1 }^\dagger b_{ - {\vec k}_2 }^\dagger | \Psi_0 \rangle
\end{align}

\section{Expectation value for number of particles per site in the Mott 
Ground State}

The expectation value for the number of particles per lattice site in
a Mott insulating phase can be easily found y noticing that
\beq \langle n_i \rangle = \langle \psi_1 | n_i | \psi_1
\rangle = \langle \psi_1 | a_i^\dagger a_i | \psi_1 \rangle \eneq
\noindent Denoting the state $|a_i \rangle \equiv a_i |\psi_1 \rangle$
\begin{align}
\nonumber |a_i \rangle &= a_i \frac{ (a_0^\dagger)^N }{ \sqrt{N!} } 
| 0 \rangle \\
a_o^\dagger &= \frac{ a_i^\dagger }{ \sqrt{N} } + \frac{1}{\sqrt{N}} 
\sum_{i = 2}^N a_i^\dagger 
\end{align}
\noindent and taking advantage of the commutation relations
\begin{align}
[a_i, a_0^\dagger] &= \frac{1}{\sqrt{N}} \\
\nonumber [a_i, (a_0^\dagger)^N] &= [a_i, a_0^\dagger (a_0^\dagger)^{N-1}] \\
\nonumber &= a_i a_0^\dagger (a_0^\dagger)^{N-1} - a_0^\dagger 
(a_0^\dagger)^{N-1}a_i \\
\nonumber &= (\frac{1}{\sqrt{N}} + a_0^\dagger a_i) (a_0^\dagger)^{N-1} 
- a_0^\dagger (a_0^\dagger)^{N-1} a_i \\
\nonumber &= a_0^\dagger [a_i, (a_0^\dagger)^{N-1}] 
+ \frac{1}{\sqrt{N}} (a_0^\dagger)^{N-1} \\
\nonumber &= a_0^\dagger ( a_i a_0^\dagger (a_0^\dagger)^{N-2} 
- a_0^\dagger (a_0^\dagger)^{N-2} a_i ) \\
\nonumber & \qquad \qquad + \frac{1}{\sqrt{N}} (a_0^\dagger)^{N-1} \\
\nonumber &= a_0^\dagger ( (\frac{1}{\sqrt{N}} + a_0^\dagger a_i) 
- a_0^\dagger (a_0^\dagger)^{N-2} a_i ) \\
\nonumber & \qquad \qquad + \frac{1}{\sqrt{N}} (a_0^\dagger)^{N-1} \\
\nonumber &= \frac{ 2 (a_0^\dagger)^{N-1} }{ \sqrt{N} } 
+ (a_0^\dagger)^2 [a_i, (a_0^\dagger)^{N-2}] \\
[a_i, (a_0^\dagger)^N] &= \sqrt{N} (a_0^\dagger)^{N-1}
\end{align}
\noindent we finally find
\begin{align}
\nonumber |a_i \rangle &= \left( \frac{ a_i (a_0^\dagger)^N }{ \sqrt{N!} } 
- \frac{ (a_0^\dagger)^N a_i }{ \sqrt{N!} } \right ) | 0 \rangle \\
\nonumber |a_i \rangle &= \frac{ [a_i, (a_0^\dagger)^N] }{ \sqrt{N!} } 
| 0 \rangle \\
&= \frac{\sqrt{N}}{\sqrt{N!}} (a_0^\dagger)^{N-1} | 0 \rangle 
= \frac{1}{\sqrt{(N-1)!}} (a_0^\dagger)^{N-1} | 0 \rangle 
\end{align}
\beqs
\langle a_i | a_i \rangle = \langle n_i \rangle = \frac{1}{\sqrt{(N-1)!}} 
\langle 0 | a_0^{N-1} (a_0^\dagger)^{N-1} | 0 \rangle
\eneqs
\beq
\langle a_i | a_i \rangle = \langle n_i \rangle = 1
\eneq
\noindent On the other hand
\beqs
\langle n_i^2 \rangle = \langle \psi_1 | n_i^2 | \psi_1 \rangle 
\eneqs
\begin{align}
\nonumber | n_i \rangle \equiv n_i | \psi_1 \rangle &= a_i^\dagger a_i | 
\psi_1 \rangle \\
\nonumber &= \frac{a_i^\dagger a_i}{\sqrt{N!}} (a_0^\dagger)^N | 0 \rangle 
= \frac{a_i^\dagger}{\sqrt{N!}} [a_i, (a_0^\dagger)^N] | 0 \rangle \\
&= \frac{a_i^\dagger}{\sqrt{N!}} \sqrt{N} (a_0^\dagger)^{N-1} 
| 0 \rangle \\
\nonumber \langle n_i | n_i \rangle &= \frac{N}{N!} \langle 0 | a_0^{N-1} a_i 
a_i^\dagger (a_0^\dagger)^{N-1}| 0 \rangle \\
\nonumber &= \frac{1}{(N-1)!} \langle 0 | a_0^{N-1} ( 1 + a_i^\dagger a_i ) 
(a_0^\dagger)^{N-1}| 0 \rangle \\
\nonumber &= \frac{1}{(N-1)!} ( (N-1)! + \langle 0 | a_0^{N-1} a_i^\dagger a_i 
(a_0^\dagger)^{N-1}| 0 \rangle \\
\nonumber &= 1 + \langle 0 | ( \sqrt{N-1} \; a_0^{N-2} + a_i^\dagger 
(a_0^\dagger)^{N-2} ) \\
\nonumber & \qquad \qquad \times ( \sqrt{N-1} \; (a_0^\dagger)^{N-2} 
+ (a_0^\dagger)^{N-1} a_i ) | 0 \rangle \\
\nonumber &= 1 + \frac{ (N-1) (N-2)! }{ (N-1)! } \\
\langle n_i | n_i \rangle &= 1 + \frac{ (N-1)! }{ (N-1)! }
\end{align}
\noindent Thus
\beq
\langle n_i^2 \rangle = 2
\eneq

\section{Impossibility of Spontaneous Symmetry Breaking in Lower Dimensional 
systems}\label{1d}

In the present appendix we review how a continuous symmetry cannot be
broken even at zero temperature in one dimension. This is the quantum
version of the so called Mermin-Wagner theorem \cite{wagner} on
melting of order in one and two dimensions by thermal fluctuations. An
earlier reference with this result is Phil Anderson's lecture notes on
solids \cite{phil2} where the melting of order by quantum fluctuation
is treated too. We consider the effective low energy Hamiltonian for a
macroscopic system with a broken symmetry ground state:
\beq
\mathcal{H} = \sum_{\vec k} \frac{\omega_{\vec k}}{2} ( b_{\vec
k}^\dagger b_{\vec k} + b_{\vec k} b_{\vec k}^\dagger )
\eneq
\noindent where $\omega_{\vec k}$ is the frequency, or energy, of the
excitation, and $b_{\vec k}, b_{\vec k}^\dagger$ are the annihilation
and creation operators of elementary excitations from the ground
state. Since we are supposing that this is the effective Hamiltonian
of a continuous broken symmetry ground state, Goldstone's theorem
implies that $\omega_{\vec k} \rightarrow 0$ as $\vec k \rightarrow 0$
\cite{gold}. We, of course, have in mind the Bogolyubov Hamiltonian,
but this applies to any Hamiltonian having a continuous broken
symmetry lowest energy state. 

This Hamiltonian can be written in terms of coordinates and
momenta. The coordinates would represent the density at wave vectors
$\vec k$ of the broken symmetry breaking. This is done through the
definition 
\beq
Q_{\vec k} = \frac{1}{\sqrt{2 \omega_{\vec k}}} ( b_{\vec k} +
b_{\vec k}^\dagger )
\eneq
\beq
P_{\vec k} = i \sqrt{\frac{\omega_{\vec k}}{2}} ( b_{\vec k}^\dagger -
b_{\vec k} )
\eneq
\noindent which are, of course, canonically conjugate variables, that
is, $[Q_{\vec k}, P_{\vec k}] = i\langle
Q_{\vec k}\rangle = 0$. The Hamiltonian then is easily
seen to be a collection of harmonic oscillators for each $\vec k$:
\beq
\mathcal{H} = \sum_{\vec k} \left( \frac{1}{2}P_{\vec k}^2 +
\frac{\omega_{\vec k}^2}{2} Q_{\vec k}^2 \right )
\eneq

In the ground state, the oscillators are centered on average $\langle
Q_{\vec k}\rangle = 0$. On the other hand, the uncertainty principle
provides for density fluctuations $\langle Q_{\vec k}^2\rangle \neq
0$. The magnitude of the fluctuations can be estimated readily because
the ground state energy is virialized.
\beq
\left \langle \frac{\omega_{\vec k}^2}{2} Q_{\vec k}^2 \right \rangle
= \frac{1}{2} E_{\text{ground state}} = \frac{1}{4} \omega_{\vec k}
\eneq
\noindent We thus see that $\langle Q_{\vec k}^2\rangle = 1/(2
\omega_{\vec k})$. Due to Goldstone's theorem, this quantity diverges
at long wavelengths. This need not be a problem if the density of
states vanishes at long wavelengths faster than the
divergence. Otherwise, uncontrolled quantum mechanical density
fluctuations will melt the ordered ground state and the symmetry will
not be broken. This would lead to absence of superfluidity or phase
stiffness of the ground state. In 1-D, the total fluctuations is given
by
\beq
\Delta Q^2 = \int \; dk \; \langle Q_k^2 \rangle = \int \; \frac{dk}{2
\omega_k}
\eneq
\noindent which for the Bogolyubov ground state, $\omega_k \sim k$,
diverges logarithmically at long wavelengths. Therefore, superfluidity
is not possible in one dimension. In the case that the finite size of
the system cuts off the divergence, the system will still not be a
superfluid, as this will happen when the energy spacings are
experimentally discernible, which in turn means the system is not a
even a fluid anymore for it does not behave like a continuum.

\end{document}